\documentclass[10pt,twocolumn,letterpaper]{article}

\usepackage[pagenumbers]{cvpr}      

\definecolor{cvprblue}{rgb}{0.21,0.49,0.74}
\usepackage[pagebackref,breaklinks,colorlinks,allcolors=cvprblue]{hyperref}

\usepackage{array}
\usepackage{multirow}
\usepackage{arydshln}
\usepackage{bm}
\usepackage{overpic}
\newcommand{\figref}[1]{Figure~\ref{#1}}
\newcommand{\tabref}[1]{Table~\ref{#1}}
\newcommand{\secref}[1]{Section~\ref{#1}}

\newcommand{\equref}[1]{Eq.~(\ref{#1})}

\usepackage[dvipsnames]{xcolor}

\def\paperID{3642} 
\def\confName{CVPR}
\def\confYear{2026}
\graphicspath{{fig/}}

\title{Time-Aware One Step Diffusion Network for Real-World Image Super-Resolution}

\author{Tianyi Zhang\textsuperscript{1}\quad Zheng-Peng Duan\textsuperscript{1}\quad Peng-Tao Jiang\textsuperscript{2}\quad Bo Li\textsuperscript{2}\quad \\
Ming-Ming Cheng\textsuperscript{1}\quad 
Chun-Le Guo\textsuperscript{1,3} \thanks{Corresponding author}\quad 
Chongyi Li\textsuperscript{1,3}\quad \\
\textsuperscript{1}VCIP, CS, Nankai University\quad
\textsuperscript{2}vivo Mobile Communication Co. Ltd\quad 
\textsuperscript{3}NKIARI, Shenzhen Futian \\ 
{\tt\small \{zty557,adamduan0211\}@mail.nankai.edu.cn,}
{\tt\small  \{pt.jiang, librad\}@vivo.com,}\\
{\tt\small \{cmm, guochunle, lichongyi\}@nankai.edu.cn}}

\begin{document}

\maketitle
\begin{abstract}
Diffusion-based real-world image super-resolution (Real-ISR) methods have demonstrated impressive performance. 
To achieve efficient Real-ISR, many works employ Variational Score Distillation (VSD) to distill pre-trained stable-diffusion (SD) model for one-step SR with a fixed timestep. 
However, since SD will perform different generative priors at different timesteps, a fixed timestep is difficult for these methods to fully leverage the generative priors in SD, leading to suboptimal performance.
To address this, we propose a \textbf{T}ime-\textbf{A}ware one-step \textbf{D}iffusion Network for Real-ISR (\textbf{TADSR}). 
We first introduce a Time-Aware VAE Encoder, which projects the same image into different latent features based on timesteps.
Through joint dynamic variation of timesteps and latent features, the student model can better align with the input pattern distribution of the pre-trained SD, thereby enabling more effective utilization of SD's generative capabilities.
To better activate the generative prior of SD at different timesteps, we propose a Time-Aware VSD loss that bridges the timesteps of the student model and those of the teacher model, thereby producing more consistent generative prior guidance conditioned on timesteps. 
Additionally, though utilizing the generative prior in SD at different timesteps, our method can naturally achieve \textbf{controllable trade-offs between fidelity and realism} by changing the timestep.
Experimental results demonstrate that our method achieves both state-of-the-art performance and controllable SR results with only a single step. 
The source codes are released at \url{https://github.com/zty557/TADSR}
\end{abstract}

\section{Introduction}
\label{sec:intro}
\begin{figure}[t]
  \includegraphics[width=\linewidth]{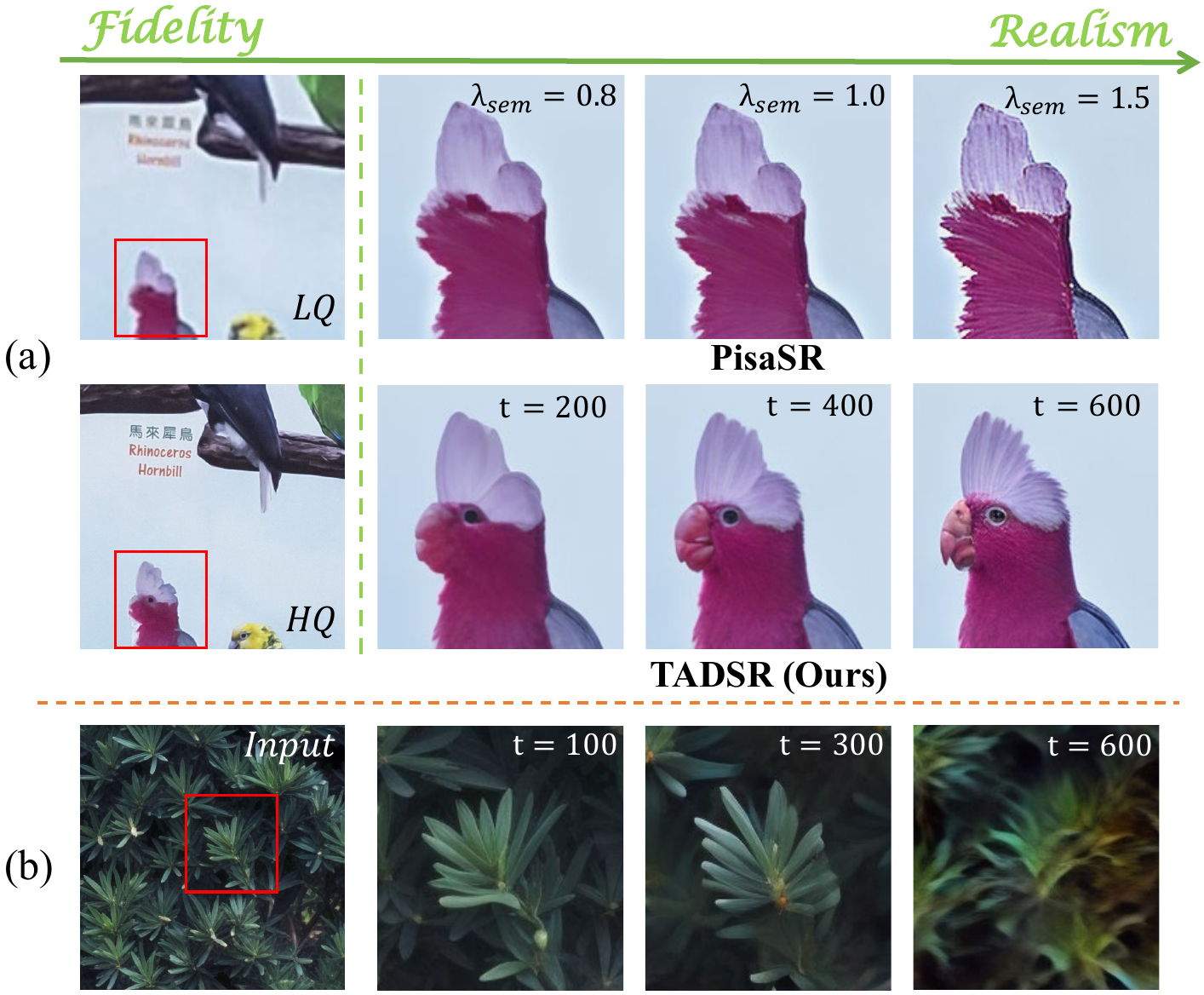}
  \caption{(a) Comparison between our TADSR(Ours) and PisaSR~\cite{sun2025pixel}. In PisaSR, increasing the semantic weight $\lambda_{sem}$ leads to restore more realistic images. As the timestep condition $t$ increases, our model recovers a more realistic parrot image. In contrast, PisaSR shows only an increase in sharpness as $\lambda_{sem}$ increases. (b) The input image and the corresponding outputs of the SD at different timesteps $t$. The outputs vary significantly across different timesteps, reflecting distinct generative priors.}
\label{fig:Teaser}
\end{figure}

Real-World Image Super-Resolution (Real-ISR) aims to restore high-quality (HQ) images from low-quality (LQ) inputs degraded by complex and unknown factors in real-world scenarios, 
which has recently attracted increasing attention~\cite{wang2021real,xie2023desra,zhang2021designing,liang2022details,duan2025dit4sr,yue2023resshift}.
To address a broader spectrum of degradation types and achieve more realistic results, many researchers have turned to generative models, particularly diffusion models~\cite{ho2020denoising}.
Consequently, several works have explored leveraging the generative priors in pre-trained Stable Diffusion (SD) models~\cite{rombach2022high} to tackle Real-ISR, yielding impressive results~\cite{wang2023exploiting,lin2023diffbir,yang2023pasd,wu2023seesr,duan2025dit4sr,wu2025omgsr}. 
Nevertheless, the iterative denoising process inherent in diffusion models introduces significant computational overhead and latency.

To overcome these limitations, some researchers have focused on distilling SD into an efficient one-step model for Real-ISR~\cite{wu2024one,sun2025pixel,chen2025adversarial,dong2025tsd,zhang2024degradation,yue2025arbitrary}. 
Specifically, OSEDiff~\cite{wu2024one} first leverages the Variational Score Distillation (VSD) loss~\cite{wang2023prolificdreamer} to distill the SD model, enabling realistic image reconstruction in a single step. 
Subsequently, S3Diff~\cite{zhang2024degradation}, PisaSR~\cite{sun2025pixel}, AdcSR~\cite{chen2025adversarial}, and TSDSR~\cite{dong2025tsd} also adopt distillation-based approaches to develop SD-based one-step Real-ISR models, using either adversarial loss or modified VSD loss. They typically use a pre-trained SD with trainable LoRA modules as the student model to perform Real-ISR, while employing a fixed-weight SD as the teacher to provide generative guidance.

However, these methods generally fix the timestep (e.g. step 999) injected into the student model while randomly sampling the timestep injected into the teacher model, which prevents them from effectively leveraging the generative prior in SD.
Specifically, as shown in Figure~\ref{fig:Teaser}(b), when the timestep $t$ equals 100, most of the image information is preserved, and the teacher’s output only differs in texture details. 
As the $t$ grows to 300, the teacher model increasingly activates leaf-related generative priors to predict content that has been obscured by noise. 
However, with $t$ increasing further to 600, most of the image information is lost, and the teacher model can only recover the overall structure and color of the leaves.
This observation indicates that the pretrained SD model exhibits different generative priors at different timesteps.
Therefore, existing methods generally suffer from the following two problems:
(1) The fixed timestep injected into the student model fails to fully leverage the generative priors at different timesteps in the pretrained SD model;
(2) The randomly sampled timestep injected into the teacher model makes it difficult to provide consistent generative guidance.
As a result, as shown in Figure \ref{fig:Teaser}(a), although we increase the semantic weight $\lambda_{sem}$ in PisaSR~\cite{sun2025pixel}, it only produces enhanced sharpness without significantly enriching the semantic content. In contrast, our method gradually generates a more realistic parrot as the timestep increases.

In the end, we propose Time-Aware One Step Diffusion Network for Super-Resolution (TADSR), a framework that more effectively distills the generative prior of SD at different timesteps into a one-step diffusion model for Real-ISR. 
To address the first limitation and better exploit the generative priors at different timesteps, two conditions need to be satisfied:
(1) the student model should receive randomly sampled timesteps;
(2) the latent features fed into the student model should vary with the timestep, reflecting the noise-level changes in SD.
Therefore, we incorporate a Time-Aware VAE Encoder (TAE), which introduces a time embedding layer into the VAE encoder to map the same image to different latent representations based on the timestep.
To address the second limitation and ensure consistent generative guidance, we propose a Time-Aware Variational Score Distillation (TAVSD) Loss, which associates the timestep injected into the student model with the one used in the teacher model through a mapping function. 
When the student model is conditioned on a larger timestep, the teacher receives a latent image corrupted with stronger noise, providing guidance that emphasizes stronger semantic generation in the reconstruction results. 
Conversely, a smaller timestep leads to similar results with reconstruction, primarily enhancing texture details.
Therefore, TAVSD can provide a more consistent generative guidance condition on the injected timestep in the student model. Our contributions are summarized as follows:

\begin{itemize}
    \item We propose \textbf{TADSR}, a \textbf{Time-Aware One-Step Diffusion Network} for Real-ISR, which naturally leverages the generative priors of SD at different timesteps to achieve controllable trade-offs between fidelity and realism in Real-ISR.
    Our TADSR achieves \textbf{superior performance} compared with other SD-based Real-ISR methods on both real-world and synthetic datasets.
    \item We propose a \textbf{Time-Aware VAE Encoder (TAE)}, which maps the same image to different latent representations based on the timestep, enabling the student model to fully exploit the generative priors at various timesteps.
    \item We propose a \textbf{Time-Aware Variational Score Distillation (TAVSD) Loss}, which aligns the timestep of the student and teacher models, providing consistent generative guidance at different timesteps.
\end{itemize}

\section{Related Work}

\subsection{Real-World Image Super-Resolution}
Traditional Image Super-Resolution (ISR) methods~\cite{edsr,rcan,chen2021pre,liang2021swinir,chen2023activating,dong2014learning,zhao2025systematic} typically degrade HQ images using simple downsampling operations to construct HQ-LQ image pairs for training.
However, these approaches struggle to handle images degraded by complex real-world processes. 
To better simulate the unknown and complex degradations in real-world scenarios, several studies~\cite{zhang2021designing,wang2021real} have proposed more sophisticated degradation pipelines to synthesize LQ data. 
Specifically, BSRGAN~\cite{zhang2021designing} introduces a random combination of basic degradation operations (e.g., downsampling, blurring, noise) injection, with varying intensities to generate realistic HQ-LQ pairs. 
Real-ESRGAN~\cite{wang2021real} proposes a second-order degradation scheme to cover a broader range of degradation types. 
In addition, inspired by Generative Adversarial Networks (GANs), researchers have adopted adversarial losses to encourage the reconstruction of more realistic images. 
Although these GAN-based methods can produce richer texture details compared to traditional approaches, they are often unstable to train and prone to generating unnatural artifacts~\cite{wu2024one}.

\begin{figure*}[t]
  \includegraphics[width=\linewidth]{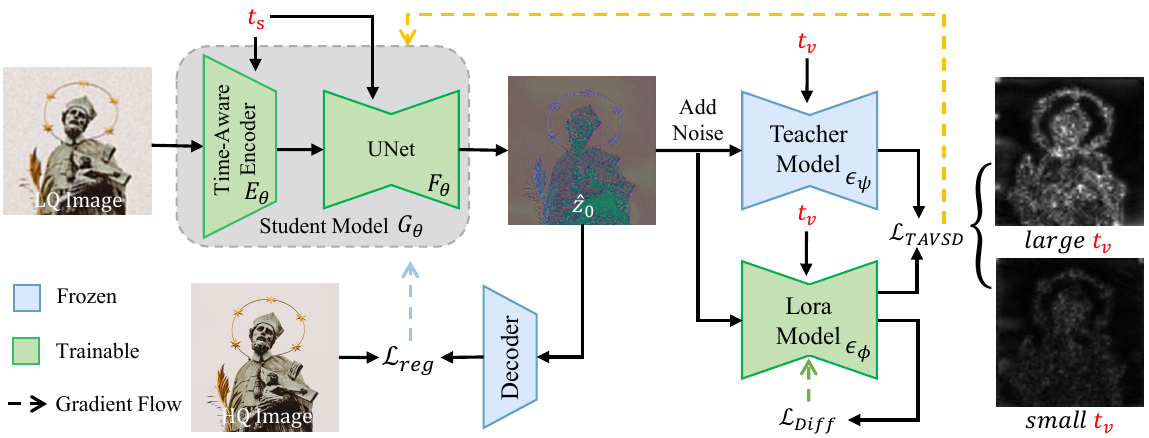}
  \caption{Overview of TADSR. We train a Student Model $G_\theta$ to perform one-step Real-ISR, which consists of a Time-Aware VAE Encoder $E_\theta$ and a UNet $F_\theta$. We randomly sample a timestep $t_s$ and map it to $t_v$. The $t_s$ and the LQ image are fed into the encoder $E_\theta$ to obtain the LQ latent. Then, $t_s$ and the LQ latent are fed into the UNet $F_\theta$ to produce the reconstructed latent feature $\hat{z}_0$. After adding noise to $\hat{z}_0$ corresponding to $t_v$, we feed it and $t_v$ into the teacher model and the LoRA model to compute the TAVSD loss (\textcolor[RGB]{255,192,0}{orange flow}). The reconstruction loss (\textcolor[RGB]{157,195,230}{blue flow}) in pixel space and TAVSD loss is then used to jointly update the student model $G_\theta$. For the LoRA Model, we employ the diffusion loss (\textcolor[RGB]{122,173,71}{green flow}) for training.}
\label{fig:overview}
\end{figure*}

\subsection{Diffusion-Based Real-ISR}
Recently, many researchers have leveraged the powerful generative priors of pre-trained diffusion models for Real-ISR tasks to achieve realistic image reconstruction. 
For example, StableSR~\cite{wang2023exploiting} conditions the diffusion process on LQ images by injecting them through a learnable time-aware encoder into the SD model, enabling strong detail generation capabilities. 
DiffBIR~\cite{lin2023diffbir} utilizes ControlNet to extract structural information from LQ images to better guide the generative prior of SD for SR. 
PASD~\cite{yang2023pasd} and SeeSR~\cite{wu2023seesr} extract semantic information from LQ inputs and inject it into SD, resulting in more realistic outputs. Although these approaches yield impressive results, the multi-step denoising process leads to high computational and time costs. 
To accelerate diffusion-based Real-ISR, OSEDiff introduces the VSD loss to distill the pre-trained SD model, enabling realistic image reconstruction in a single step. 
S3Diff further adopts degradation-guided LoRA adapters combined with adversarial training to achieve one-step SR. 
PisaSR trains two separate LoRA adapters for pixel-level and semantic-level guidance, allowing controllable trade-offs between realism and fidelity.

However, these methods overlook the varying generative capabilities of SD in different timesteps and train with a fixed timestep. 
Our work aims to fully exploit these time-dependent generative prior to achieve superior SR performance and a natural balance between fidelity and realism.

\section{Methodology}
\subsection{Problem Definition}
\label{sec:problem}
Real-ISR aims to reconstruct HQ images $x_H$ from LQ images $x_L$ that suffer from complex and unknown degradations. 
With the advancement of deep learning, researchers have commonly adopted neural networks $G_\theta$ to estimate the HQ images and optimize the network through loss functions. 
The general form of the loss function is:

\begin{equation}
\label{eq:eq1}
\begin{gathered}
\theta^* = \arg\min_{\theta} \mathbb{E}_{(x_L, x_H) \sim \mathcal{D}} [
\mathcal{L}_{Rec}(G_\theta(x_L), x_H) \\
+ \lambda \mathcal{L}_{Reg}(G_\theta(x_L))
],    
\end{gathered}
\end{equation}
where the $\mathcal{L}_{Rec}$ denotes the reconstruction loss to optimize the fidelity of the reconstructed results. $\mathcal{L}_{Reg}$ is the regression loss to enhance the realism of the results, and $\lambda$ is a hyperparameter to balance $\mathcal{L}_{Rec}$ and $\mathcal{L}_{Reg}$.

Recently, with the advancement of diffusion models, several studies~\cite{wu2024one,sun2025pixel} have leveraged the generative prior in pre-trained SD and adopted VSD as a regression objective. 
VSD is designed to align the distribution of generated (fake) images with that of real images.
Specifically, a pretrained teacher model trained on real images is used to estimate the score of the real image distribution.
In addition, a diffusion model trained on the generator’s outputs provides an estimate of the score of the fake distribution.
The generator is then optimized by minimizing the discrepancy between these two score estimates, thereby encouraging it to produce samples that are indistinguishable from real images.
The formation of VSD is:

\begin{equation}
\label{eq:eq2}
\nabla_{\theta} \mathcal{L}_{VSD}(\hat{z}, c)
= \mathbb{E}_{t, \epsilon} \left[
\omega(t) \left( \epsilon_{\psi}(\hat{z}_t; t, c) - \epsilon_{\phi}(\hat{z}_t; t, c) \right)
\frac{\partial \hat{z}}{\partial \theta}
\right],    
\end{equation}
where  $\epsilon_{\psi}$ is the pre-trained diffusion model (teacher model) to estimate the real score, $\epsilon_{\phi}$ represents its replica with trainable LoRA (LoRA model) to predict the fake score, and $c$ is a text embedding of a caption describing the input image. 
$\hat{z}=G_\theta(x_{L})$ is the output of the student network $G_\theta$, and $\hat{z}_t=\alpha_t\hat{z}+\beta_t\epsilon$ is the noisy latent input. 
$\epsilon$ is the gaussian noise, and $\alpha_t$ and $\beta_t$ are the scale parameters in diffusion.

Formally, the VSD loss can be viewed as the residual between the noise outputs of the teacher model and the LoRA model, which can be transformed to the residual of their predicted latent images through $z=\frac{z_t-\beta_t\epsilon}{\alpha_t}$. 
In practice, the gradient of the VSD loss is applied only to the generator and does not propagate through the teacher model.
Therefore, the loss can be expressed in the following form:

\begin{equation}
\begin{gathered}
\mathcal{L}_{VSD}(\theta)
= \mathbb{E}_{t, \epsilon} [
\omega'(t) \| G_\theta(x_L)- G_{\theta^-}(x_L) \\
- z_{\psi^-}(\hat{z}_t; t, c) + z_{\phi^-}(\hat{z}_t; t, c) \|^2_2 ],   
\end{gathered}
\label{eq:eq3}
\end{equation}
where $^-$ denotes $stopgrad()$, $z_{\psi}$ and $z_{\phi}$ denote the predicted latent image of the teacher model and LoRA model, respectively. 
Therefore, we can decode the latent images predicted by the teacher model and the LoRA model into the image space to analyze the guidance provided by VSD.

\subsection{Overview}
\label{sec:overview}
As illustrated in Figure \ref{fig:overview}, we distill a Student Model $G_\theta$ to perform one-step Real-ISR, which consists of a trainable Time-Aware Encoder $E_\theta$ and a LoRA finetuned diffusion UNet $F_\theta$. 
Following the OSEDiff~\cite{wu2024one}, we use a pretrained SD model as the Teacher Model $\epsilon_\psi$ and its replica with trainable LoRA as the LoRA Model to obtain the fake score.
First, we sample an HQ-LQ image pair from the dataset and a timestep $t_s$ from a uniform distribution in the range of 0 to 999.
Then, both the LQ image and $t_{s}$ are fed into Student Model $G_\theta$ to obtain the latent output $\hat{z}_0$. 
We decode $\hat{z}_0$ into pixel space and compute the reconstruction loss with the HQ image.
In the latent space, we map the $t_s$ to another timestep $t_v$, and add noise corresponding to timestep $t_v$ to the $\hat{z}_0$ to obtain $\hat{z}_{t_v}$.
Then, we feed $\hat{z}_{t_v}$ and $t_v$ into both the teacher model and the LoRA model to compute the Time-Aware Variational Score Distillation loss $\mathcal{L}_{TAVSD}$ to enhance the realism. 
Consistent with Figure~\ref{fig:Teaser}(b), when $t_v$ is small, the gradients produced by the TAVSD loss are relatively small and mainly reflect texture details. 
In contrast, when $t_v$ is large, the gradients become significantly larger and provide more semantic guidance. 
Therefore, the TAVSD loss can offer more consistent gradient guidance condition on $t_s$, enabling better distillation of the teacher model.
In addition, the LoRA model is trained with a diffusion loss using the data generated by the student model.

\begin{figure}[t]
  \includegraphics[width=\linewidth]{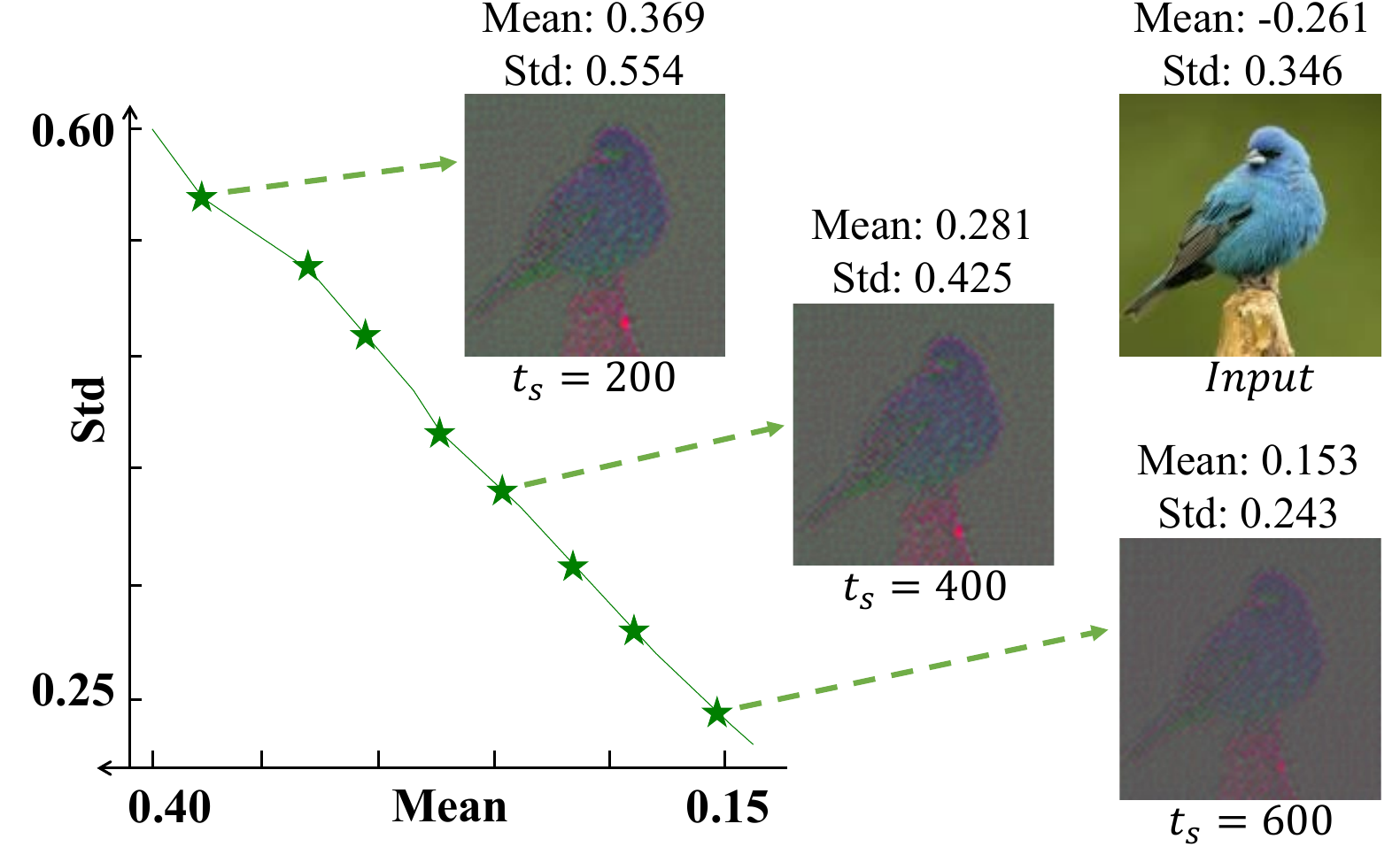}
  \caption{PCA visualization of latent features produced by TAE under different timesteps $t_s$, and the corresponding mean and standard deviation (Std) of latent features. TAE can encode the same image into distinct latent features conditioned on different timesteps, which aligns with the synchronized variation between timesteps and latent features in the pre-trained SD.}
\label{fig:Tae}
\end{figure}

\subsection{Time-Aware VAE Encoder}
In the original SD, as the timestep increases, the input latent features are injected with more noise, thereby activating different generative priors to produce the image.
Considering the importance of the timestep in the diffusion process, multi-step SD-based Real-ISR methods often take the timestep as a condition along with the LR image to control the SR process — as seen in the time-aware encoder of StableSR and the ControlNet used in SeeSR and DiffBIR.

However, since multi-step denoising iterations are not required, existing one-step diffusion-based Real-ISR methods generally overlook the role of the timestep and train with only a fixed timestep, making it difficult to fully exploit the generative priors in SD at different timesteps.
A straightforward improvement is to randomly sample timestep during training. 
However, since the original VAE encoder maps the same image to a single latent distribution regardless of the timestep, it is difficult for the diffusion network to effectively activate different generative priors based solely on the timestep, given the same latent input. 
In the original diffusion process, variations in the timestep reflect changes in the noise level of the latent distribution. 
We argue that a one-step SD-based Real-ISR network should exhibit a similar property in order to fully leverage the corresponding generative priors.
Clearly, directly injecting noise into the latent distribution according to the timestep is inappropriate, as it would compromise reconstruction fidelity.

Therefore, we propose a Time-Aware VAE Encoder (TAE) to better utilize the generative priors in SD.
By incorporating a temporal embedding layer into the VAE encoder, TAE encodes the input image into different latent distributions conditioned on the timestep $t_s$, enabling synchronized variation between $t_s$ and latent distribution, thus better activating the generative priors at different timesteps within SD. This process can be formulated as:

\begin{equation}
\label{eq:eq4}
    z_L = E_\theta(x_L, t_s), \hat{z}=F_\theta(z_L, t_s),
\end{equation}
where $E_\theta$ is the TAE model and $F_\theta$ is the Unet model.

As shown in Figure \ref{fig:Tae}, TAE encodes the same input image into different latent feature condition on the timestep $t_s$. 
Overall, as the $t_s$ increases, both the mean and variance of the latent features show a decreasing trend. 
After visualizing the latent feature via PCA dimensionality reduction, we can also clearly observe the changes in the latent space.

\subsection{Time-Aware Variational Score Distillation}
Following the OSEDiff~\cite{wu2024one}, the VSD loss has been widely adopted in SD-based one-step Real-ISR methods to enhance the realism of reconstruction results. 
In one-step image generation tasks, the timestep is usually sampled randomly to distill the full generative prior of the teacher model. 
However, we find that in Real-ISR, this paradigm instead leads to \textbf{inconsistent generative guidance}, since SD exhibits different generative priors across timesteps, while the timestep sampling in the teacher model is completely random and independent of the student model. 

\begin{figure}[t]
  \includegraphics[width=\linewidth]{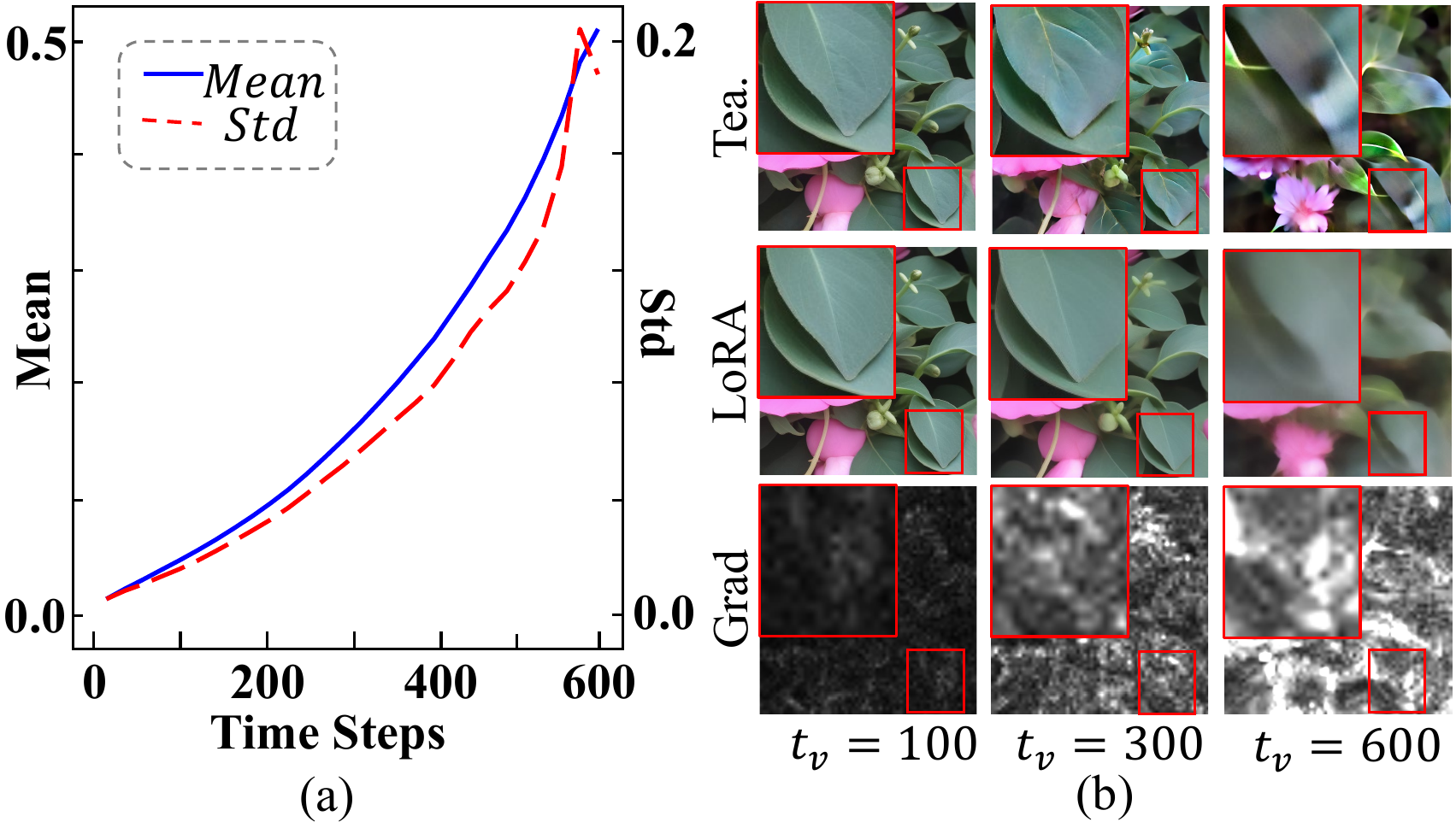}
  \caption{(a) Mean and standard deviation (Std) of the VSD loss at different timesteps. (b) The outputs of the teacher model and the LoRA model are decoded into pixel space and gradients in latent space at different timesteps $t$.}
\label{fig:Tavsd}
\end{figure}

As discussed in \secref{sec:problem}, due to the stop-gradient operation, the guidance of the VSD loss can be interpreted as the residual between the latent images predicted by the teacher model and the LoRA model. 
Therefore, we decode these latent images into the pixel space and analyze the guidance of the VSD loss at different timesteps. 
As illustrated in Figure~\ref{fig:Tavsd}, first, the mean and standard deviation of the VSD loss exhibit a clear upward trend as the timestep increases. Besides, we observe that at $t_v$ equals 100, the outputs of the teacher and LoRA models are similar, and the gradients mainly reflect enhancements in texture details.
In contrast, when $t_v$ increases to 300, the teacher model’s output contains significantly more semantic information while the LoRA model’s output remains smooth, and the gradients reflect global semantic guidance.
However, when $t_v$ increases to 600, the teacher model can only recover coarse color and structural information from the noisy latent input, making it difficult to provide meaningful guidance. 
This implies that the VSD loss provides distinct guidance for the same image depending on $t_v$. 
Such opposing directional guidance creates conflicting optimization signals for the student model, leading to suboptimal convergence.

\begin{table*}[t]
\fontsize{12pt}{16pt}\selectfont
\centering
\caption{A comprehensive evaluation against state-of-the-art methods across synthetic and real-world datasets. The top-performing and runner-up results under each metric are marked in \textcolor{red}{\textbf{red}} and \textcolor{blue}{\underline{blue}}, respectively.}
\resizebox{\textwidth}{!}{
\begin{tabular}{@{}c | c | c c c c c c c c c c @{}}
\hline
Datasets &
Metrics &
StableSR &
DiffBIR &
SeeSR &
SinSR &
S3Diff &
OSEDiff &
PisaSR &
TSDSR &
AdcSR &
TADSR \\
\hline
\multirow{9}{*}{\begin{tabular}[c]{@{}c@{}}
\textit{DIV2k-Val}\end{tabular}}
& PSNR $\uparrow$ &23.261 &23.409 &23.679 &\textcolor{red}{\textbf{24.417}} &23.530 &23.723 &\textcolor{blue}{\underline{23.867}} & 22.17 &23.743 &23.815  \\
& SSIM $\uparrow$ &0.5726 &0.5732 &0.6043 &0.6023 &0.5933 &\textcolor{red}{\textbf{0.6109}} &\textcolor{blue}{\underline{0.6058}} & 0.5602 &0.6017 &0.6028  \\
& LPIPS $\downarrow$ &0.3113 &0.3456 &0.3194 &0.3235 &\textcolor{red}{\textbf{0.2581}} &0.2942 &0.2823 &\textcolor{blue}{\underline{0.2736}} &0.2853 &0.3078  \\
& CLIPIQA $\uparrow$ &0.6771 &0.7082 &0.6935 &0.6505 &0.7001 &0.6682 &0.6928 &\textcolor{blue}{\underline{0.7149}} &0.6763 &\textcolor{red}{\textbf{0.7353}} \\
& MUSIQ $\uparrow$ &65.918 &68.396 &68.665 &62.838 &67.923 &67.971 &\textcolor{blue}{\underline{69.681}} &\textcolor{red}{\textbf{70.65}} &67.995 &69.649 \\
& MANIQA $\uparrow$ &0.6174 &0.6355 &0.6222 &0.5392 &0.6311 &0.6132 &\textcolor{blue}{\underline{0.6375}} &0.6077 &0.6073 &\textcolor{red}{\textbf{0.6443}}  \\
& TOPIQ $\uparrow$ &0.5979 &0.6344 &\textcolor{blue}{\underline{0.6856}} &0.5721 &0.6334 &0.6188 &0.6619 &0.6672 &0.6526 &\textcolor{red}{\textbf{0.7044}}  \\
& QALIGN $\uparrow$ &3.5273 &3.8774 &\textcolor{blue}{\underline{3.9765}} &3.5159 &3.8666 &3.8357 &3.8812 &3.927 &3.612 &\textcolor{red}{\textbf{4.0783}} \\

\hline

\multirow{9}{*}{\begin{tabular}[c]{@{}c@{}}
\textit{DRealSR}\end{tabular}}

& PSNR $\uparrow$ &28.030 &25.929 &28.073 &\textcolor{blue}{\underline{28.345}} &27.539 &27.915 &28.318 &26.20 &28.099 &\textcolor{red}{\textbf{28.387}}\\
& SSIM $\uparrow$ &0.7536 &0.6526 &0.7684 &0.7491 &0.7491 &\textcolor{red}{\textbf{0.7833}} &\textcolor{blue}{\underline{0.7804}} &0.7170 &0.7726 &0.7758 \\
& LPIPS $\downarrow$ &0.3284 &0.4518 &0.3173 &0.3697 &0.3109 &\textcolor{blue}{\underline{0.2968}} &\textcolor{red}{\textbf{0.2960}} &0.3116 &0.3046 &0.3235 \\
& CLIPIQA $\uparrow$ &0.6356 &0.6863 &0.6909 &0.6375 &0.7131 &0.6974 &0.6971  &\textcolor{blue}{\underline{0.7309}} &0.7049 &\textcolor{red}{\textbf{0.7398}}  \\
& MUSIQ $\uparrow$ &58.512 &65.667 &65.080 &55.009 &63.941 &64.691 &66.113 &66.12 &\textcolor{blue}{\underline{66.266}} &\textcolor{red}{\textbf{67.016}}  \\
& MANIQA $\uparrow$ &0.5594 &\textcolor{blue}{\underline{0.6289}} &0.6051 &0.4894 &0.6124 &0.5903 &0.6160 &0.5820 &0.5915 &\textcolor{red}{\textbf{0.6309}}  \\
& TOPIQ $\uparrow$ &0.5323 &0.6149 &\textcolor{blue}{\underline{0.6575}} &0.5122 &0.6040 &0.6002 &0.6333 &0.6251 &0.6527 &\textcolor{red}{\textbf{0.6758}}  \\
& QALIGN $\uparrow$ &3.0614 &3.6011 &3.5882 &3.0982 &3.6148 &3.5450 &3.5838 &3.6928 &\textcolor{blue}{\underline{3.6520}} &\textcolor{red}{\textbf{3.7491}}  \\

\hline

\multirow{9}{*}{\begin{tabular}[c]{@{}c@{}}
\textit{RealSR}\end{tabular}}

& PSNR $\uparrow$ &24.645 &24.240 &25.149 &\textcolor{red}{\textbf{26.266}} &25.183 &25.148 &\textcolor{blue}{\underline{25.503}} &23.404 &25.469 &25.166  \\
& SSIM $\uparrow$ &0.7080 &0.6649 &0.7211 &\textcolor{blue}{\underline{0.7341}} &0.7269 &0.7338 &\textcolor{red}{\textbf{0.7418}} &0.6886 &0.7301 &0.7150 \\
& LPIPS $\downarrow$ &0.3002 &0.3470 &0.3007 &0.3241 &\textcolor{blue}{\underline{0.2721}} &0.2920 &\textcolor{red}{\textbf{0.2672}} &0.2805 &0.2885 &0.3168 \\
& CLIPIQA $\uparrow$ &0.6234 &0.6959 &0.6699 &0.6153 &0.6731 &0.6687 &0.6699 &\textcolor{blue}{\underline{0.7199}} &0.6731 &\textcolor{red}{\textbf{0.7283}} \\
& MUSIQ $\uparrow$ &65.883 &68.340 &69.819 &60.575 &65.824 &69.087 &70.147 &\textcolor{blue}{\underline{70.7710}} &69.899 &\textcolor{red}{\textbf{71.182}} \\
& MANIQA $\uparrow$ &0.6230 &0.6530 &0.6450 &0.5409 &0.6427 &0.6337 &\textcolor{blue}{\underline{0.6551}} &0.6311 &0.6353 &\textcolor{red}{\textbf{0.6715}} \\
& TOPIQ $\uparrow$ &0.5748 &0.6052 &\textcolor{blue}{\underline{0.6890}} &0.5188 &0.6162 &0.6251 &0.6374 &0.6642 &0.6793 &\textcolor{red}{\textbf{0.7082}}  \\
& QALIGN $\uparrow$ &3.2767 &3.6313 &3.7190 &3.1889 &3.6638 &3.6915 &3.6355 &3.7748 &\textcolor{blue}{\underline{3.7749}} &\textcolor{red}{\textbf{3.9477}}  \\

\hline


\multirow{5}{*}{\begin{tabular}[c]{@{}c@{}}
\textit{RealLR200}\end{tabular}}

& CLIPIQA $\uparrow$ &0.6036 &0.7072 &0.7023 &0.6474 &0.7122 &0.6792 &0.7153 &\textcolor{blue}{\underline{0.7248}} &0.7048 &\textcolor{red}{\textbf{0.7741}}  \\
& MUSIQ $\uparrow$ &62.863 &67.727 &70.195 &63.126 &68.897 &69.041 &\textcolor{blue}{\underline{70.935}} &70.930 &69.759  &\textcolor{red}{\textbf{72.166}}\\
& MANIQA $\uparrow$ &0.5922 &0.6464 &0.6482 &0.5522 &0.6536 &0.6331 &\textcolor{blue}{\underline{0.6639}} &0.6363 &0.6354 &\textcolor{red}{\textbf{0.6738}}  \\
& TOPIQ $\uparrow$ &0.5286 &0.5905 &\textcolor{blue}{\underline{0.6900}} &0.5689 &0.6401 &0.5990 &0.6627 &0.6664 &0.6684 &\textcolor{red}{\textbf{0.7249}}  \\
& QALIGN $\uparrow$ &3.3409 &3.7782 &\textcolor{blue}{\underline{4.0305}} &3.4105 &3.9228 &3.8459 &3.9891 &3.8908 &3.9312 &\textcolor{red}{\textbf{4.2622}}  \\

\hline
\end{tabular}
}
\label{tab:compare_with_other_methods}
\vspace{-2mm}
\end{table*}

This phenomenon arises because most of the image information is preserved when $t$ is small, and both the teacher and LoRA models focus mainly on generating texture details. 
As $t$ increases, the injection of more noise gradually obscures the underlying image content, forcing the teacher model to rely more heavily on its generative prior. 
However, since the LoRA model is trained on low-quality data generated by the student model and does not employ the CFG strategy~\cite{ho2022classifier}, its outputs tend to be overly smooth.
This also explains why the VSD loss can make images more realistic while avoiding over-smoothing — since the LoRA model tends to produce smoother outputs than the teacher model, their residuals naturally provide high-frequency details that guide the generation process.

Considering that the guidance provided by VSD varies across different timesteps, we establish a connection between the timestep $t_s$ input to the student model and the timestep $t_v$ in the teacher model, so that the VSD loss can provide more consistent gradient guidance conditioned on $t_s$. 
Specifically, we feed the randomly sampled $t_s$ and the LQ image into the student model to obtain $\hat{z}=G_\theta(x_L, t_s)$. Then, $t_s$ is mapped to $t_v$ by:
\begin{equation}
\label{eq:eq5}
    t_{v}=\lambda t_{s} + \gamma, t_s\in[0,999], t_v\in[0,999]
\end{equation}
where the $\lambda$ and $\gamma$ are the hyperparameters.

We then add noise to $\hat{z}$ corresponding to $t_v$ to obtain $\hat{z}_{t_{v}} = \alpha_{t_v} \hat{z}+\beta_{t_v}\epsilon$, which is fed to the teacher model and the LoRA model with $t_v$ to compute the TAVSD loss:

\begin{equation}
\label{eq:eq6}
\begin{gathered}
    \nabla_{\theta} \mathcal{L}_{TAVSD}(\hat{z}, c, t_v)
= \mathbb{E}_{\epsilon} [
\omega(t_{v}) ( \epsilon_{\psi}(\hat{z}_{t_{v}}; t_{v}, c) \\
- \epsilon_{\phi}(\hat{z}_{t_{v}}; t_{v}, c) )
\frac{\partial \hat{z}}{\partial \theta}
]. 
\end{gathered}
\end{equation}

By leveraging the TAVSD loss, the model can naturally balance generation and fidelity in the Real-ISR task simply by varying the timestep condition $t_s$ input to the model.

\begin{figure*}[t]
  \includegraphics[width=\linewidth]{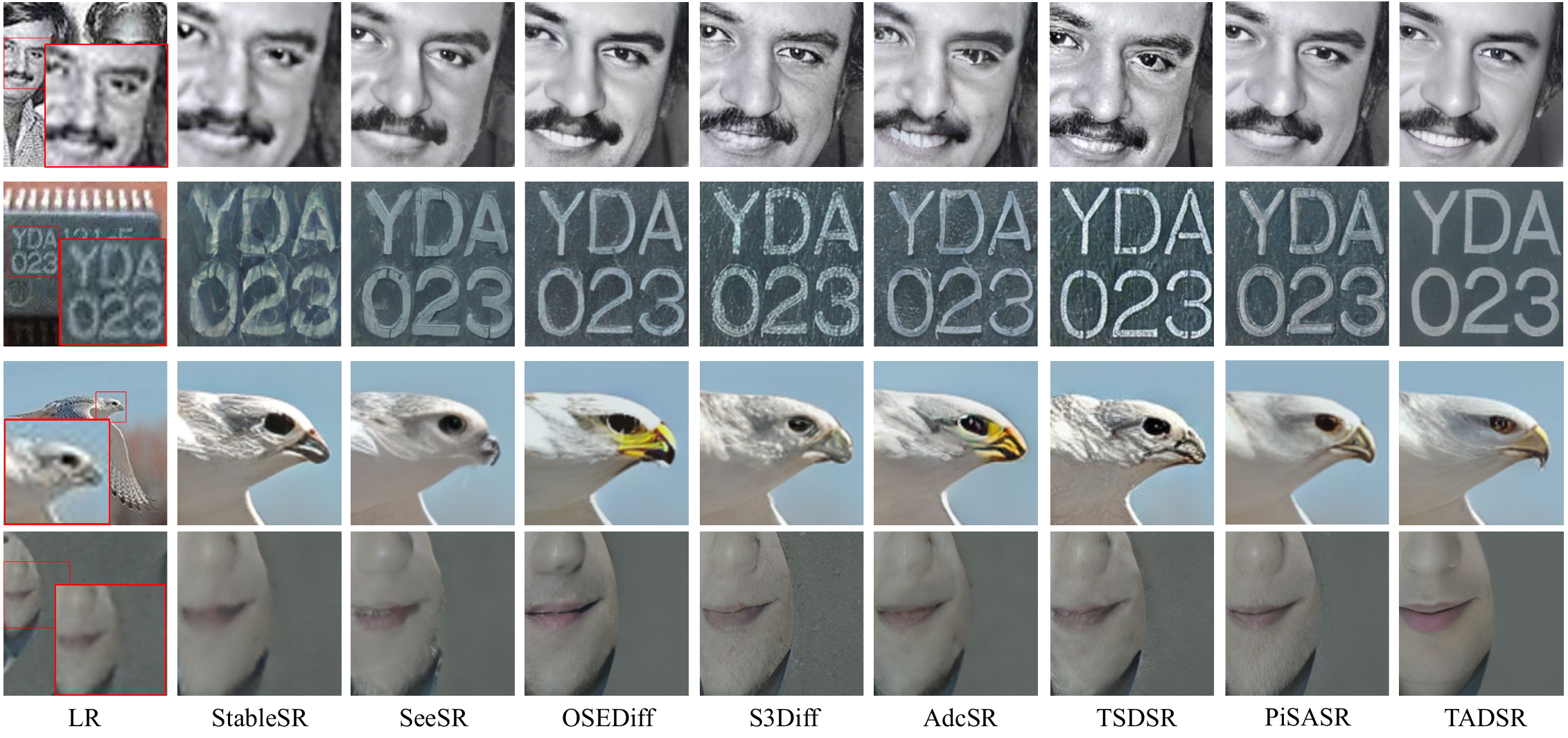}
  \caption{Visual comparisons between our method and other Real-ISR methods. Please zoom in for a better view.}
\label{fig:visionCompare}
\end{figure*}

\subsection{Training Loss}
We train the student model with reconstruction and regression losses.
To avoid gradient inconsistency arising from the ill-posed problem of the Real-ISR~\cite{liang2022details} while fully leveraging the teacher model knowledge, we first apply a Gaussian blur to both the reconstructed image and the HQ image $x_H$ before computing the MSE loss.
This ensures that the $x_H$ only supervises the low-frequency content of the reconstruction, helping to preserve high-frequency details. We adopt a larger blur kernel for larger timesteps $t_s$, which enhances the trade-off between fidelity and generation:
\begin{equation}
\label{eq:eq7}
    \mathcal{L}_{MSE}^{blur} = \mathcal{L}_{MSE} \left( G_\theta(x_L)*G_{t_s}, x_H*G_{t_s} \right).
\end{equation}
Where $*$ denotes the convolution operation, $G_{t_s}$ is the convolution kernel whose size is determined by $t_s$.
This loss and the LPIPS loss form the reconstruction loss:
\begin{equation}
\label{eq:eq8}
    \mathcal{L}_{Rec} = \mathcal{L}_{MSE}^{blur} + \mathcal{L}_{LPIPS}(G_\theta(x_L), x_H).
\end{equation}

For the regression loss, we adopt the TAVSD loss in \equref{eq:eq6} to improve the realism of the generated results. The overall loss for the student model is:
\begin{equation}
\label{eq:eq9}
    \mathcal{L}_{Stu} = \mathcal{L}_{Rec} + \lambda_{TAVSD} \cdot \mathcal{L}_{TAVSD}.
\end{equation}

We adopt the original diffusion loss for the LoRA model:
\begin{equation}
\label{eq:eq10}
    \mathcal{L}_{Diff}(\hat{\mathbf{z}}, c_y) = \mathbb{E}_{t, \epsilon} \left[ \left\| \epsilon_{\phi}(\hat{\mathbf{z}}_t; t, c_y) - \epsilon' \right\|^2 \right],
\end{equation}
where $\epsilon'$ is the randomly sampled Gaussian noise as the training target for the denoising network.

\section{Experiments}
\subsection{Experimental Setup}
\textbf{Training.}
We use LSDIR~\cite{li2023lsdir} as the training data with the $512 \times 512$ patch size. 
To generate paired HQ-LQ training data, we follow the degradation pipeline from Real-ESRGAN~\cite{wang2021real}. 
We use AdamW optimizer~\cite{loshchilov2017decoupled} with a learning rate $5 \times 10^{-5}$ and set LoRA rank to 4 for both the student model and LoRA model.
We employ the SD 2.1-base as the pre-trained  model and fine-tune it for 2k iterations using 8 NVIDIA A40 GPUs with a batch size of 24.
For the choice of hyperparameters, we set $\lambda=0.5$ and $\gamma=0$ in \equref{eq:eq5}.
We adopt the degradation-aware prompt extraction (DAPE) module~\cite{wu2023seesr} to extract text prompts.

\noindent
\textbf{Test Dataset.}
We evaluate our method in both synthetic and real-world dataset. 
For the synthetic dataset, we randomly crop 3K patches with a resolution of 512 × 512 from the DIV2K~\cite{div2k} validation set and synthesize LQ data using the same pipeline as that in training. 
For real-world data, we employ RealSR~\cite{realsr}, DrealSR~\cite{drealsr}, and RealLR200~\cite{wu2023seesr}. 
We center-crop RealSR~\cite{realsr} and DrealSR~\cite{drealsr} datasets with size $128 \times 128$ for LQ images and $512 \times 512$ for HQ image. 
For RealLR200~\cite{wu2023seesr} dataset, since the corresponding HQ images are unavailable, we perform only a $128 \times 128$ center-crop on the LQ images.

\noindent
\textbf{Evaluation Metrics.} 
We utilize several reference and non-reference metrics to evaluate the performance of various methods on the test data.
For the reference measures, we employ PSNR, SSIM~\cite{wang2004image}, and LPIPS~\cite{lpips} to measure image fidelity. 
For the non-reference measures, we employ CLIPIQA~\cite{clipiqa}, MUSIQ~(\cite{ke2021musiq}, MANIQA~\cite{yang2022maniqa}, TOPIQ~\cite{chen2024topiq}, and QALIGN~\cite{qalign-pmlr-v235-wu24ah} to measure image quality.

\noindent
\textbf{Compared Methods.}
We compare our method with several multi-step diffusion-based methods StableSR~\cite{wang2023exploiting}, DiffBIR~\cite{lin2023diffbir}, SeeSR~\cite{wu2023seesr}, and one-step methods SinSR~\cite{wang2023sinsr}, OSEDiff~\cite{wu2024one}, S3Diff~\cite{zhang2024degradation}, AdcSR~\cite{chen2025adversarial}, 
TSDSR~\cite{dong2025tsd}, and PisaSR~\cite{sun2025pixel}. 
All comparative results are obtained using publicly released code and model weights for testing.

\subsection{Comparisons with State-of-the-art Methods}

\noindent
\textbf{Quantitative Comparisons.} We set up the timestep condition $t_s=500$ in our method, and show the quantitative comparisons on the four synthetic and real-world datasets in \tabref{tab:compare_with_other_methods}. We have the following observations:
(1) TADSR achieves the highest no-reference scores across four datasets, except for the MUSIQ on DIV2K-Val.
This demonstrates that TADSR can more effectively leverage the generative priors from SD to produce more realistic results. 
Notably, TADSR is the only one-step method that consistently outperforms multi-step methods on all no-reference metrics, achieving both efficiency and perceptual quality.
(2) TADSR maintains PSNR values comparable to other SD-based one-step methods, indicating a good balance between fidelity and realism.
(3) TADSR shows clear improvements over other SD-based one-step methods on CLIPIQA and TOPIQ, highlighting its superior semantic awareness and generative capability.
(4) Under the same number of parameters as OSEDiff, TADSR significantly outperforms OSEDiff on no-reference metrics while maintaining comparable reference metrics, further demonstrating TADSR’s effective utilization of generative priors and the resulting performance improvements.

\noindent
\textbf{Qualitative Comparisons.} 
Figure \ref{fig:visionCompare} shows the visual comparisons between our method and the other state-of-the-art Real-ISR methods. 
As shown in the first row, TADSR generates significantly more natural and sharper textures from heavily degraded LQ images, especially in facial regions such as the teeth, eyes, and eyebrows, demonstrating its strong semantic generation capability.
In the second row, the digits and letters produced by TADSR appear much clearer, showcasing its superior degradation removal ability while preserving fidelity.
In the third row, TADSR yields more natural results around the eagle's eyes and beak.
In the fourth row, only TADSR accurately restored natural-looking facial features such as the nose, mouth, and chin. 
Other methods generally suffered from degradation, resulting in some distortion, and failed to reconstruct a plausible chin structure.
Overall, thanks to the ability to distill generative priors from SD more effectively in TAVSD loss, TADSR can produce natural and realistic results in a single diffusion step. 
Compared to other methods, it achieves strong perceptual quality while maintaining high efficiency.

\subsection{Ablation Study}

\begin{figure}[t]
  \includegraphics[width=\linewidth]{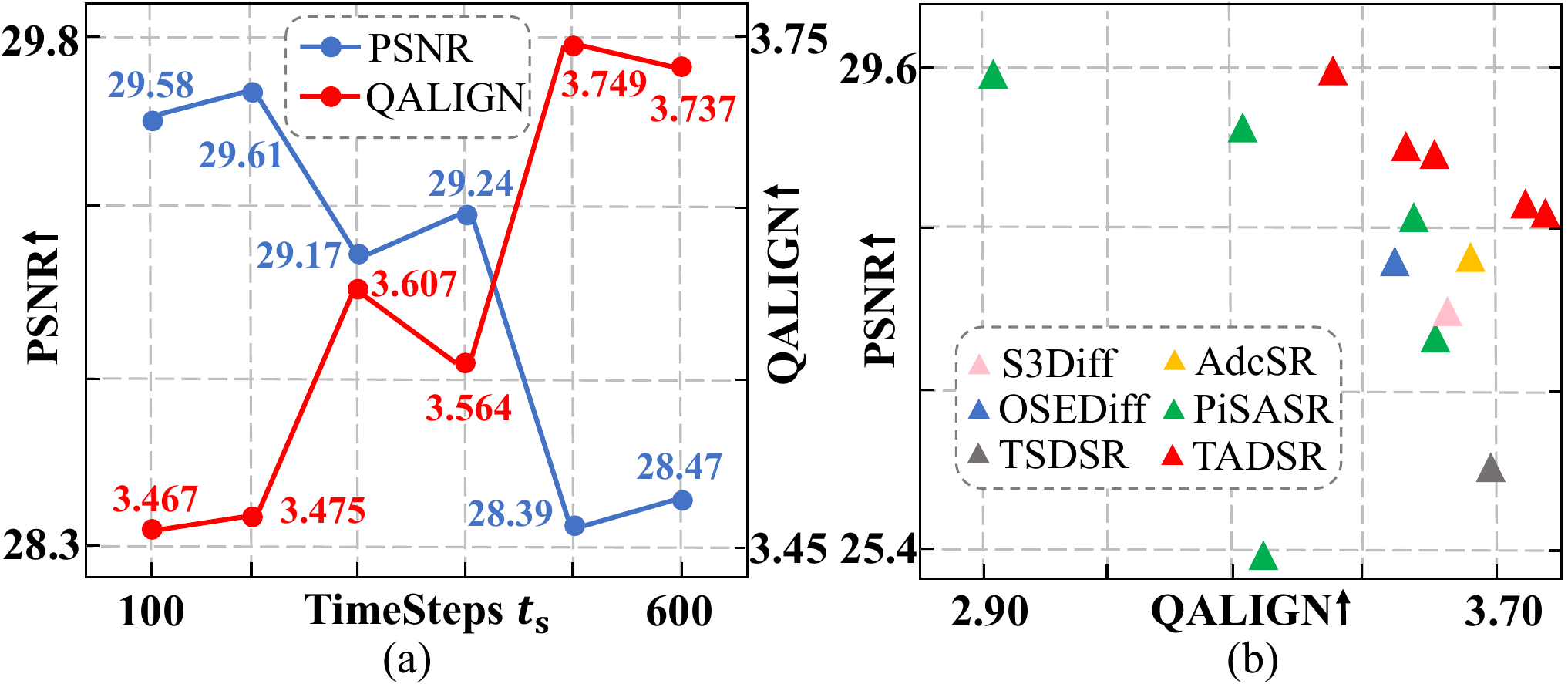}
  \caption{(a) Quantitative metrics of our method under different timestep $t_s$, evaluated on the \textit{DrealSR} dataset. (b) Comparison of our method under different timesteps $t_s$, PisaSR under different semantic guidance weights $\lambda_{sem}$, and other one-step diffusion-based Real-ISR methods, evaluated on the \textit{DrealSR} dataset.}
\label{fig:tsCompare}
\end{figure}

\noindent
\textbf{Impact of Different Timestep Condition.}
As shown in Figure~\ref{fig:tsCompare}(a), we analyze the impact of timestep $t_s$ in our method on both reference and no-reference metrics. 
As $t_s$ increases, PSNR exhibits a decreasing trend while QALIGN shows an upward trend, indicating a trade-off where fidelity is sacrificed to enhance realism.
This trade-off between fidelity and realism aligns with the function of $t_s$, as a larger $t_s$ means that TAVSD provides stronger generative guidance, while a smaller $t_s$ provides more fidelity-preserving guidance. 
Similar visual results can be observed in \textbf{supplementary materials}.
Furthermore, we compare the results of our method under different $t_s$, PisaSR under different 
$\lambda_{sem}$ settings, and other one-step Real-ISR methods, as shown in Figure \ref{fig:tsCompare}(b). 
It can be observed that our method consistently lies in the top-right corner across different $t_s$. 
When $t_s$ equals 200, our method achieves 26.61dB PSNR, which is more than 1dB higher than SinSR, and QALIGN is significantly higher than SinSR. 
In contrast, although PisaSR can also achieve a PSNR of 29.60dB by tuning the $\lambda_{pix}=1.0$ and $\lambda_{sem}=0.6$, its QALIGN is only 2.91, which is similar to SinSR.
This indicates that our method achieves a substantial improvement in fidelity with only a minimal compromise in realism.

\begin{figure}[t]
  \includegraphics[width=\linewidth]{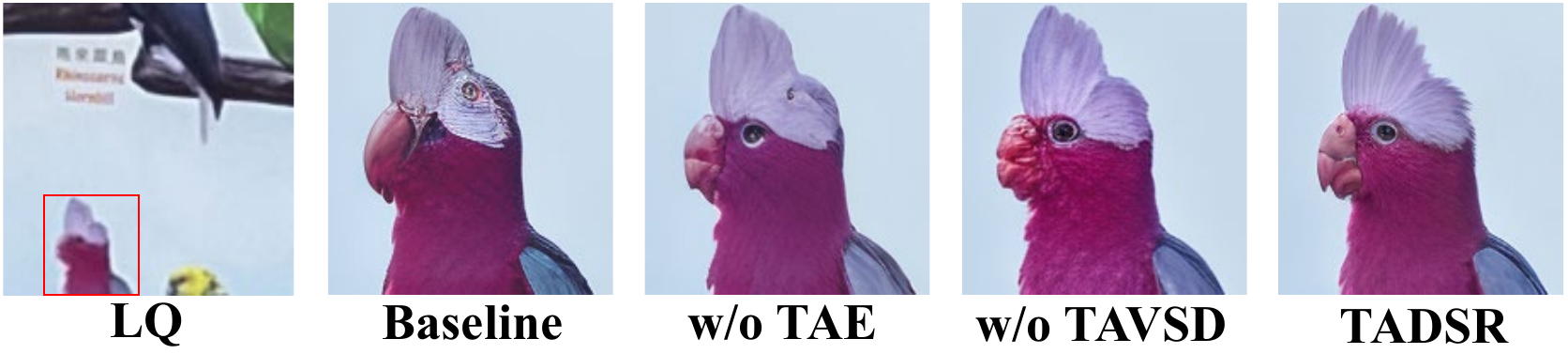}
  \caption{Vision Comparisons of the ablation study on TAE and TAVSD. Baseline use the original VAE encoder and VSD loss.}
\label{fig:ablation}
\end{figure}

\begin{table}[t]
  \small
  \centering
  \setlength\tabcolsep{3.2pt}
    \begin{tabular}{c|c|c|c|c|c} \hline
    Dataset & Method & PSNR$\uparrow$ & MUSIQ$\uparrow$ & CLIPIQA $\uparrow$ & TOPIQ$\uparrow$ \\ \hline
    \multirow{4}[0]{*}{RealSR} & Baseline & 24.39 & 70.22 & 0.6751 & 0.6391 \\ 
     & w/o TAE & \textcolor{blue}{\underline{24.89}} & 70.08 & 0.6857 & 0.6466 \\
     & w/o TAVSD & 24.84 & \textcolor{blue}{\underline{70.96}} & \textcolor{blue}{\underline{0.6930}} & \textcolor{blue}{\underline{0.6553}} \\ 
     & Full & \textcolor{red}{\textbf{25.16}} & \textcolor{red}{\textbf{71.18}} & \textcolor{red}{\textbf{0.7283}} & \textcolor{red}{\textbf{0.7082}} \\ \hline
    \multirow{4}[0]{*}{DrealSR} & Baseline & 27.45 & 65.90 & 0.6887 & 0.6275 \\ 
     & w/o TAE & 27.95 & 65.95 & \textcolor{blue}{\underline{0.7030}} & \textcolor{blue}{\underline{0.6396}} \\ 
     & w/o TAVSD & \textcolor{blue}{\underline{28.03}} & \textcolor{blue}{\underline{66.95}} & 0.7015 & 0.6373 \\ 
     & Full & \textcolor{red}{\textbf{28.39}} & \textcolor{red}{\textbf{67.02}} & \textcolor{red}{\textbf{0.7398}} & \textcolor{red}{\textbf{0.6758}} \\ \hline
    \end{tabular}%
    \caption{Quantitative Comparison of ablation study on TAVSD and TAE. Baseline uses the original VAE encoder and VSD loss.}
  \label{tab:ablation}
  \vspace{-0.5em}
\end{table}%

\noindent
\textbf{Impact of TAVSD and TAE.}
To validate the effectiveness of TAVSD and TAE, we conducted ablation studies by removing them.
We employ the original VAE encoder in SD and the VSD loss as our baseline and conduct ablation studies by separately removing TAE and TAVSD. 
We use PSNR to evaluate fidelity and CLIPIQA, MUSIQ, and TOPIQ to assess realism. 
As shown in \tabref{tab:ablation}, we have the following three key observations: 
(1) Although the baseline also adopts randomly sampled timesteps during training, after removing TAE, both reference and no-reference metrics decline, demonstrating that timestep-adaptive latent distribution plays a crucial role in effectively utilizing the generative priors in SD.
(2) When TAVSD is ablated, all metrics similarly decrease, indicating that more consistent guidance from the teacher model better activates generative priors across different timesteps.
(3) Baseline shows significant degradation in PSNR and moderate decline in others, proving that both TAE and TAVSD improve fidelity and realism.
Additionally, Figure~\ref{fig:ablation} presents a visual comparison of our ablation studies, showing that both the absence of TAE/TAVSD leads to unrealistic parrot reconstructions, while the baseline even produces visible artifacts.
In contrast, our method produces realistic and natural results by fully exploiting the generative priors in SD.

\section{Conclusion}
In this paper, we propose TADSR, a one-step SD-based Real-ISR method.
TADSR introduces a variable timestep $t_s$ into the student model and uses a Time-Aware VAE Encoder to fully utilize the generative priors in SD at different timesteps. 
To further distill the priors at different timesteps in SD to achieve varied SR effects, TADSR leverages the Time-Aware Variational Score Distillation to provide more consistent generative guidance condition on $t_s$.
As a result, TADSR fully leverages the generative priors in SD and naturally achieves a controllable trade-off between fidelity and realism condition on $t_s$.
Our experiments demonstrate that TADSR achieves state-of-the-art performance among all Real-ISR methods.

{
\small
\bibliographystyle{ieeenat_fullname}
\bibliography{references}

@String(CVPR= {IEEE Conf. Comput. Vis. Pattern Recog.})

@String(ICCV= {Int. Conf. Comput. Vis.})

@String(ECCV= {Eur. Conf. Comput. Vis.})

@String(NIPS= {Adv. Neural Inform. Process. Syst.})

@String(TIP  = {IEEE Trans. Image Process.})

@String(AAAI = {AAAI})

@String(CVPRW= {IEEE Conf. Comput. Vis. Pattern Recog. Worksh.})

@String(CVPR  = {CVPR})

@String(ICCV  = {ICCV})

@String(ECCV  = {ECCV})

@String(NIPS  = {NeurIPS})

@String(TIP   = {IEEE TIP})

@String(CVPRW= {CVPRW})

@String(ICML = {ICML})

@article{wu2024one,
  title={One-Step Effective Diffusion Network for Real-World Image Super-Resolution},
  author={Wu, Rongyuan and Sun, Lingchen and Ma, Zhiyuan and Zhang, Lei},
  journal={arXiv preprint arXiv:2406.08177},
  year={2024}
}

@article{yue2023resshift,
  title={Resshift: Efficient diffusion model for image super-resolution by residual shifting},
  author={Yue, Zongsheng and Wang, Jianyi and Loy, Chen Change},
  journal={arXiv preprint arXiv:2307.12348},
  year={2023}
}

@inproceedings{dong2014learning,
  title={Learning a deep convolutional network for image super-resolution},
  author={Dong, Chao and Loy, Chen Change and He, Kaiming and Tang, Xiaoou},
  booktitle=ECCV,
  pages={184--199},
  year={2014},
  organization={Springer}
}

@inproceedings{liang2021swinir,
  title={Swinir: Image restoration using swin transformer},
  author={Liang, Jingyun and Cao, Jiezhang and Sun, Guolei and Zhang, Kai and Van Gool, Luc and Timofte, Radu},
  booktitle=ICCV,
  pages={1833--1844},
  year={2021}
}

@inproceedings{chen2021pre,
  title={Pre-trained image processing transformer},
  author={Chen, Hanting and Wang, Yunhe and Guo, Tianyu and Xu, Chang and Deng, Yiping and Liu, Zhenhua and Ma, Siwei and Xu, Chunjing and Xu, Chao and Gao, Wen},
  booktitle=CVPR,
  pages={12299--12310},
  year={2021}
}

@inproceedings{chen2023activating,
  title={Activating more pixels in image super-resolution transformer},
  author={Chen, Xiangyu and Wang, Xintao and Zhou, Jiantao and Qiao, Yu and Dong, Chao},
  booktitle=CVPR,
  pages={22367--22377},
  year={2023}
}

@inproceedings{zhang2021designing,
  title={Designing a practical degradation model for deep blind image super-resolution},
  author={Zhang, Kai and Liang, Jingyun and Van Gool, Luc and Timofte, Radu},
  booktitle=ICCV,
  pages={4791--4800},
  year={2021}
}

@inproceedings{wang2021real,
  title={Real-esrgan: Training real-world blind super-resolution with pure synthetic data},
  author={Wang, Xintao and Xie, Liangbin and Dong, Chao and Shan, Ying},
  booktitle=ICCV,
  pages={1905--1914},
  year={2021}
}

@inproceedings{liang2022details,
  title={Details or artifacts: A locally discriminative learning approach to realistic image super-resolution},
  author={Liang, Jie and Zeng, Hui and Zhang, Lei},
  booktitle=CVPR,
  pages={5657--5666},
  year={2022}
}

@article{xie2023desra,
  title={Desra: detect and delete the artifacts of gan-based real-world super-resolution models},
  author={Xie, Liangbin and Wang, Xintao and Chen, Xiangyu and Li, Gen and Shan, Ying and Zhou, Jiantao and Dong, Chao},
  journal={arXiv preprint arXiv:2307.02457},
  year={2023}
}

@article{ho2020denoising,
  title={Denoising diffusion probabilistic models},
  author={Ho, Jonathan and Jain, Ajay and Abbeel, Pieter},
  journal=NIPS,
  volume={33},
  pages={6840--6851},
  year={2020}
}

@article{wang2023exploiting,
  title={Exploiting Diffusion Prior for Real-World Image Super-Resolution},
  author={Wang, Jianyi and Yue, Zongsheng and Zhou, Shangchen and Chan, Kelvin CK and Loy, Chen Change},
  journal={arXiv preprint arXiv:2305.07015},
  year={2023}
}

@article{lin2023diffbir,
  title={DiffBIR: Towards Blind Image Restoration with Generative Diffusion Prior},
  author={Lin, Xinqi and He, Jingwen and Chen, Ziyan and Lyu, Zhaoyang and Fei, Ben and Dai, Bo and Ouyang, Wanli and Qiao, Yu and Dong, Chao},
  journal={arXiv preprint arXiv:2308.15070},
  year={2023}
}

@inproceedings{div2k,
  title={Ntire 2017 challenge on single image super-resolution: Dataset and study},
  author={Agustsson, Eirikur and Timofte, Radu},
  booktitle=CVPRW,
  pages={126--135},
  year={2017}
}

@inproceedings{lpips,
  title={The unreasonable effectiveness of deep features as a perceptual metric},
  author={Zhang, Richard and Isola, Phillip and Efros, Alexei A and Shechtman, Eli and Wang, Oliver},
  booktitle=CVPR,
  pages={586--595},
  year={2018}
}

@inproceedings{clipiqa,
  title={Exploring clip for assessing the look and feel of images},
  author={Wang, Jianyi and Chan, Kelvin CK and Loy, Chen Change},
  booktitle=AAAI,
  volume={37},
  number={2},
  pages={2555--2563},
  year={2023}
}

@article{ssim,
  title={Image quality assessment: from error visibility to structural similarity},
  author={Wang, Zhou and Bovik, Alan C and Sheikh, Hamid R and Simoncelli, Eero P},
  journal=TIP,
  volume={13},
  number={4},
  pages={600--612},
  year={2004},
  publisher={IEEE}
}

@article{lora,
  title={Lora: Low-rank adaptation of large language models},
  author={Hu, Edward J and Shen, Yelong and Wallis, Phillip and Allen-Zhu, Zeyuan and Li, Yuanzhi and Wang, Shean and Wang, Lu and Chen, Weizhu},
  journal={arXiv preprint arXiv:2106.09685},
  year={2021}
}

@article{wang2023prolificdreamer,
  title={ProlificDreamer: High-Fidelity and Diverse Text-to-3D Generation with Variational Score Distillation},
  author={Wang, Zhengyi and Lu, Cheng and Wang, Yikai and Bao, Fan and Li, Chongxuan and Su, Hang and Zhu, Jun},
  journal={arXiv preprint arXiv:2305.16213},
  year={2023}
}

@inproceedings{realsr,
  title={Toward real-world single image super-resolution: A new benchmark and a new model},
  author={Cai, Jianrui and Zeng, Hui and Yong, Hongwei and Cao, Zisheng and Zhang, Lei},
  booktitle=ICCV,
  pages={3086--3095},
  year={2019}
}

@inproceedings{drealsr,
  title={Component divide-and-conquer for real-world image super-resolution},
  author={Wei, Pengxu and Xie, Ziwei and Lu, Hannan and Zhan, Zongyuan and Ye, Qixiang and Zuo, Wangmeng and Lin, Liang},
  booktitle=ECCV,
  pages={101--117},
  year={2020},
  organization={Springer}
}

@inproceedings{edsr,
  title={Enhanced deep residual networks for single image super-resolution},
  author={Lim, Bee and Son, Sanghyun and Kim, Heewon and Nah, Seungjun and Mu Lee, Kyoung},
  booktitle=CVPRW,
  pages={136--144},
  year={2017}
}

@inproceedings{rcan,
  title={Image super-resolution using very deep residual channel attention networks},
  author={Zhang, Yulun and Li, Kunpeng and Li, Kai and Wang, Lichen and Zhong, Bineng and Fu, Yun},
  booktitle=ECCV,
  pages={286--301},
  year={2018}
}

@InProceedings{qalign-pmlr-v235-wu24ah,
  title = 	 {Q-Align: Teaching {LMM}s for Visual Scoring via Discrete Text-Defined Levels},
  author =       {Wu, Haoning and Zhang, Zicheng and Zhang, Weixia and Chen, Chaofeng and Liao, Liang and Li, Chunyi and Gao, Yixuan and Wang, Annan and Zhang, Erli and Sun, Wenxiu and Yan, Qiong and Min, Xiongkuo and Zhai, Guangtao and Lin, Weisi},
  booktitle=ICML,
  pages = 	 {54015--54029},
  year = 	 {2024},
  editor = 	 {Salakhutdinov, Ruslan and Kolter, Zico and Heller, Katherine and Weller, Adrian and Oliver, Nuria and Scarlett, Jonathan and Berkenkamp, Felix},
  volume = 	 {235},
  series = 	 {Proceedings of Machine Learning Research},
  month = 	 {21--27 Jul},
  publisher =    {PMLR},
  url = 	 {https://proceedings.mlr.press/v235/wu24ah.html}
}

@inproceedings{yang2023pasd,
    title={Pixel-Aware Stable Diffusion for Realistic Image Super-Resolution and Personalized Stylization},
    author={Yang Tao and Wu Rongyuan and Ren Peiran and Xie Xuansong and Zhang Lei},
    booktitle=ECCV,
    year={2023}
}

@inproceedings{rombach2022high,
  title={High-resolution image synthesis with latent diffusion models},
  author={Rombach, Robin and Blattmann, Andreas and Lorenz, Dominik and Esser, Patrick and Ommer, Bj{\"o}rn},
  booktitle=CVPR,
  pages={10684--10695},
  year={2022}
}

@inproceedings{ke2021musiq,
  title={Musiq: Multi-scale image quality transformer},
  author={Ke, Junjie and Wang, Qifei and Wang, Yilin and Milanfar, Peyman and Yang, Feng},
  booktitle=ICCV,
  pages={5148--5157},
  year={2021}
}

@inproceedings{yang2022maniqa,
  title={Maniqa: Multi-dimension attention network for no-reference image quality assessment},
  author={Yang, Sidi and Wu, Tianhe and Shi, Shuwei and Lao, Shanshan and Gong, Yuan and Cao, Mingdeng and Wang, Jiahao and Yang, Yujiu},
  booktitle=CVPR,
  pages={1191--1200},
  year={2022}
}

@article{wang2004image,
  title={Image quality assessment: from error visibility to structural similarity},
  author={Wang, Zhou and Bovik, Alan C and Sheikh, Hamid R and Simoncelli, Eero P},
  journal=TIP,
  volume={13},
  number={4},
  pages={600--612},
  year={2004},
  publisher={IEEE}
}

@inproceedings{wu2023seesr,
  title={SeeSR: Towards Semantics-Aware Real-World Image Super-Resolution},
  author={Wu, Rongyuan and Yang, Tao and Sun, Lingchen and Zhang, Zhengqiang and Li, Shuai and Zhang, Lei},
  booktitle=CVPR,
  year={2024}
}

@inproceedings{wang2023sinsr,
  title={SinSR: Diffusion-Based Image Super-Resolution in a Single Step},
  author={Wang, Yufei and Yang, Wenhan and Chen, Xinyuan and Wang, Yaohui and Guo, Lanqing and Chau, Lap-Pui and Liu, Ziwei and Qiao, Yu and Kot, Alex C and Wen, Bihan},
  booktitle=CVPR,
  year={2024}
}

@ARTICLE{chen2024topiq,
  author={Chen, Chaofeng and Mo, Jiadi and Hou, Jingwen and Wu, Haoning and Liao, Liang and Sun, Wenxiu and Yan, Qiong and Lin, Weisi},
  journal=TIP, 
  title={TOPIQ: A Top-Down Approach From Semantics to Distortions for Image Quality Assessment}, 
  year={2024},
  volume={33},
  number={},
  pages={2404-2418}}

@inproceedings{sun2025pixel,
  title={Pixel-level and semantic-level adjustable super-resolution: A dual-lora approach},
  author={Sun, Lingchen and Wu, Rongyuan and Ma, Zhiyuan and Liu, Shuaizheng and Yi, Qiaosi and Zhang, Lei},
  booktitle={Proceedings of the Computer Vision and Pattern Recognition Conference},
  pages={2333--2343},
  year={2025}
}

@inproceedings{chen2025adversarial,
  title={Adversarial diffusion compression for real-world image super-resolution},
  author={Chen, Bin and Li, Gehui and Wu, Rongyuan and Zhang, Xindong and Chen, Jie and Zhang, Jian and Zhang, Lei},
  booktitle={Proceedings of the Computer Vision and Pattern Recognition Conference},
  pages={28208--28220},
  year={2025}
}

@inproceedings{dong2025tsd,
  title={Tsd-sr: One-step diffusion with target score distillation for real-world image super-resolution},
  author={Dong, Linwei and Fan, Qingnan and Guo, Yihong and Wang, Zhonghao and Zhang, Qi and Chen, Jinwei and Luo, Yawei and Zou, Changqing},
  booktitle={Proceedings of the Computer Vision and Pattern Recognition Conference},
  pages={23174--23184},
  year={2025}
}

@article{zhang2024degradation,
  title={Degradation-guided one-step image super-resolution with diffusion priors},
  author={Zhang, Aiping and Yue, Zongsheng and Pei, Renjing and Ren, Wenqi and Cao, Xiaochun},
  journal={arXiv preprint arXiv:2409.17058},
  year={2024}
}

@inproceedings{li2023lsdir,
  title={Lsdir: A large scale dataset for image restoration},
  author={Li, Yawei and Zhang, Kai and Liang, Jingyun and Cao, Jiezhang and Liu, Ce and Gong, Rui and Zhang, Yulun and Tang, Hao and Liu, Yun and Demandolx, Denis and others},
  booktitle={Proceedings of the IEEE/CVF Conference on Computer Vision and Pattern Recognition},
  pages={1775--1787},
  year={2023}
}

@article{loshchilov2017decoupled,
  title={Decoupled weight decay regularization},
  author={Loshchilov, Ilya and Hutter, Frank},
  journal={arXiv preprint arXiv:1711.05101},
  year={2017}
}

@article{ho2022classifier,
  title={Classifier-free diffusion guidance},
  author={Ho, Jonathan and Salimans, Tim},
  journal={arXiv preprint arXiv:2207.12598},
  year={2022}
}

@article{duan2025dit4sr,
  title={DiT4SR: Taming Diffusion Transformer for Real-World Image Super-Resolution},
  author={Duan, Zheng-Peng and Zhang, Jiawei and Jin, Xin and Zhang, Ziheng and Xiong, Zheng and Zou, Dongqing and Ren, Jimmy and Guo, Chun-Le and Li, Chongyi},
  journal={arXiv preprint arXiv:2503.23580},
  year={2025}
}

@article{zhao2025systematic,
  title={A Systematic Investigation on Deep Learning-Based Omnidirectional Image and Video Super-Resolution},
  author={Zhao, Qianqian and Guo, Chunle and Zhang, Tianyi and Zhang, Junpei and Jia, Peiyang and Su, Tan and Jiang, Wenjie and Li, Chongyi},
  journal={arXiv preprint arXiv:2506.06710},
  year={2025}
}

@inproceedings{yue2025arbitrary,
  title={Arbitrary-steps image super-resolution via diffusion inversion},
  author={Yue, Zongsheng and Liao, Kang and Loy, Chen Change},
  booktitle={Proceedings of the Computer Vision and Pattern Recognition Conference},
  pages={23153--23163},
  year={2025}
}

@article{wu2025omgsr,
  title={OMGSR: You Only Need One Mid-timestep Guidance for Real-World Image Super-Resolution},
  author={Wu, Zhiqiang and Sun, Zhaomang and Zhou, Tong and Fu, Bingtao and Cong, Ji and Dong, Yitong and Zhang, Huaqi and Tang, Xuan and Chen, Mingsong and Wei, Xian},
  journal={arXiv preprint arXiv:2508.08227},
  year={2025}
}
}

\clearpage
\maketitlesupplementary

In this supplementary material, we provide the following content:
\begin{itemize}
  \item Detailed derivation about the Variational Score Distillation loss in Section~\ref{sec:derivation}
  \item Visual comparisons and quantitative metrics of TADSR across different timesteps in Section~\ref{sec:ablation_ts}
  \item Ablation study on the blurred MSE loss in Section~\ref{sec:ablation_mse}
  \item Ablation study on the hyperparameters of TAVSD loss in Section~\ref{sec:ablation_hyper}
  \item Efficiency comparison between TADSR and other diffusion-based Real-ISR methods in Section~\ref{sec:efficiency}
  \item Comparisons with GAN-based Real-ISR methods in Section~\ref{sec:compare_gan}
  \item Extended visual comparisons with SD-based Real-ISR approaches in Section~\ref{sec:vc_sd}
\end{itemize}
 
\section{Detailed Derivation}
\label{sec:derivation}
According to the original diffusion process in SD~\cite{rombach2022high}, at step $t$, the current state $z_t$ satisfies:
\begin{equation}
\label{eq:eq11}
\boldsymbol{z}_t = \alpha_t \boldsymbol{z}_0 + \beta_t \boldsymbol{\epsilon}, t=1, 2, \dots, T,
\end{equation}
where $\alpha_t$ and $\beta_t$ are the scale parameters in diffusion, $\boldsymbol{\epsilon} \sim \mathcal{N}(\boldsymbol{0}, \boldsymbol{I}^2)$ and $\boldsymbol{z}_0$ is HR latent in Real-ISR task.
Therefore, we can express $z_0$ in terms of $z_t$ and $\epsilon$ as $\boldsymbol{z}_0 = \frac{\boldsymbol{z}_t - \beta_t \boldsymbol{\epsilon}}{\alpha_t}$.
Then, we can rewrite Eq.~(2) in the main paper as follows:
\begin{equation}
\label{eq:eq12}
\begin{aligned}
    \nabla_{\theta} \mathcal{L}_{VSD}
    &= \mathbb{E}_{t, \epsilon} \left[
    \omega(t) \left( \epsilon_{\psi}(\hat{z}_t; t, c) - \epsilon_{\phi}(\hat{z}_t; t, c) \right)
    \frac{\partial \hat{z}}{\partial \theta}
    \right],\\
    &= \mathbb{E}_{t, \epsilon} \left[
    \omega(t) \frac{\alpha_t}{\beta_t} \left( \hat{z}_{\phi}(\hat{z}_t; t, c) - \hat{z}_{\psi}(\hat{z}_t; t, c) \right)
    \frac{\partial \hat{z}}{\partial \theta}
    \right], \\
    &= \mathbb{E}_{t, \epsilon} \left[
    \omega'(t) \left( \hat{z}_{\phi}(\hat{z}_t; t, c) - \hat{z}_{\psi}(\hat{z}_t; t, c) \right)
    \frac{\partial \hat{z}}{\partial \theta}
    \right],
\end{aligned}
\end{equation}
where $\epsilon_{\psi}$ is the pre-trained diffusion model (teacher model), $\epsilon_{\phi}$ represents its replica with trainable LoRA~\cite{lora} (LoRA model), $\hat{z}_\psi$ and $\hat{z}_\phi$  represent the latent images predicted by the teacher model and the LoRA model respectively, $c$ is a text embedding of a caption describing the input image, and $\omega_t$ is a time-varying weighting function. 
Therefore, we can represent the VSD~\cite{wang2023prolificdreamer} loss using the residual between the latent images predicted by the teacher model and the LoRA model, which is then decoded into pixel space to analyze the timestep-dependent guidance.

\begin{figure*}[ht!]
  \centering
    \includegraphics[width=\linewidth]{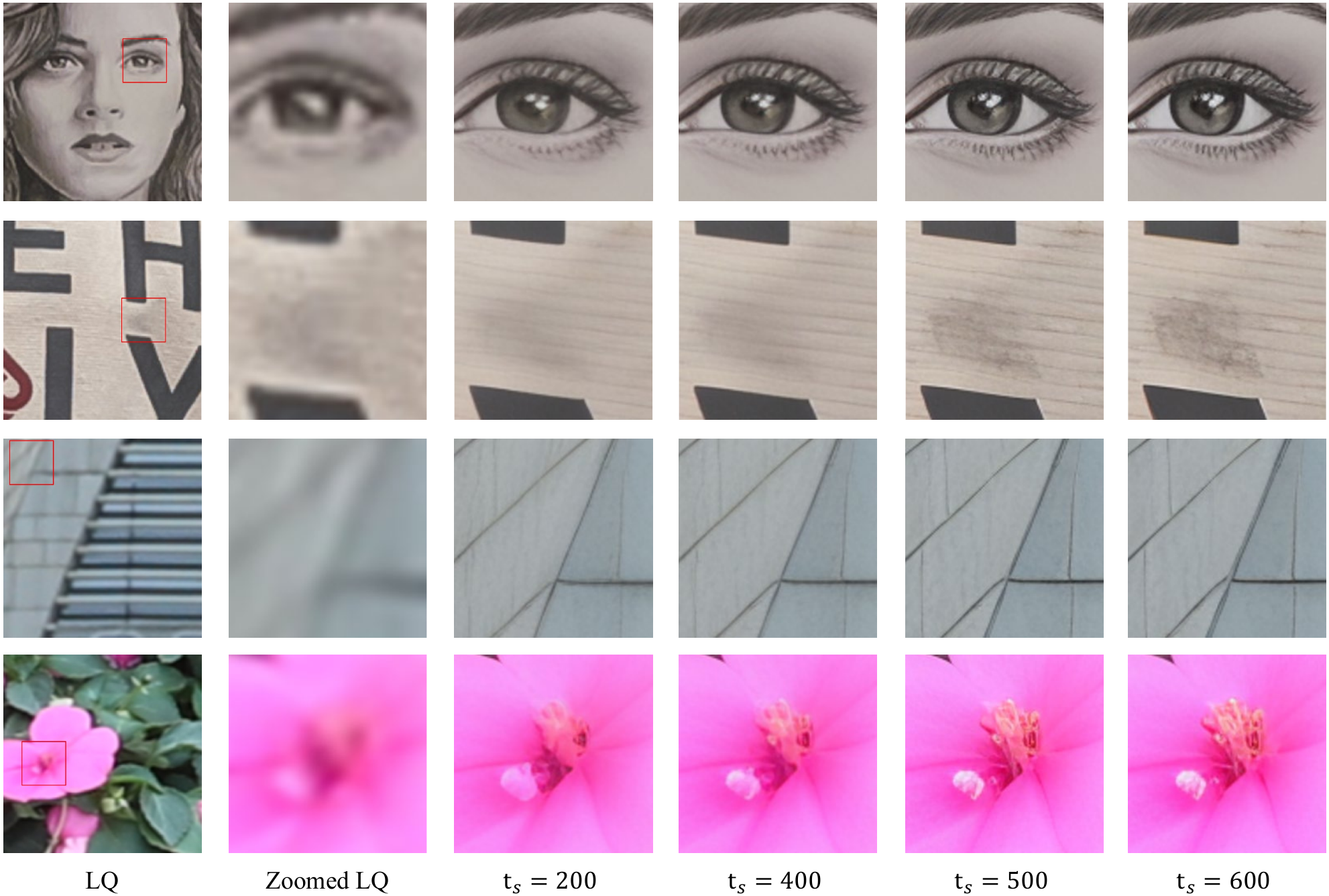}
  \caption{Vision comparisons of TADSR at different timesteps $t_s$. Zoom in for a better view.}
\label{fig:supp_vc_ts}
\end{figure*}

\begin{table}[t]
  \small
  \centering
  \caption{Quantitative comparison of ablation study on blurred MSE loss, evaluated on DrealSR~\cite{drealsr} dataset.}
  \setlength\tabcolsep{3.2pt}
    \begin{tabular}{c|c|c|c|c|c} \hline

Methods &
PSNR $\uparrow$ &
SSIM $\uparrow$  &
MUSIQ $\uparrow$ &
TOPIQ $\uparrow$ &
QALIGN $\uparrow$ \\
\hline

$t_s=100$ &29.576 &0.8021 &64.022 & 0.6201 &3.4666  \\
$t_s=200$ &\textcolor{red}{\textbf{29.610}} &\textcolor{red}{\textbf{0.8059}} &63.735 & 0.6108 &3.4750  \\
$t_s=300$ &29.167 &0.7935 &65.367 & 0.6463 &3.6069  \\
$t_s=400$ &29.245 &0.7992 &64.681 & 0.6149 &3.5635  \\
$t_s=500$ &28.387 &0.7758 &\textcolor{red}{\textbf{67.016}} & \textcolor{red}{\textbf{0.6758}} &\textcolor{red}{\textbf{3.7491}}  \\
$t_s=600$ &28.473 &0.7834 &66.620 & 0.6597 &3.7372  \\
\hline
\end{tabular}
\label{tab:ablation_timestep}
\vspace{-1em}
\end{table}

\begin{table}[t]
  \small
  \centering
  \caption{Quantitative comparison of ablation study on blurred MSE loss, evaluated on DrealSR~\cite{drealsr} dataset.}
  \setlength\tabcolsep{3.2pt}
    \begin{tabular}{c|c|c|c|c} \hline

Methods &
PSNR $\uparrow$ &
SSIM $\uparrow$  &
MUSIQ $\uparrow$ &
QALIGN $\uparrow$ \\
\hline

w/o blurred MSE &29.074 &0.7841 &64.732 &3.5299  \\
TADSR ($t_s=300$) &\textcolor{red}{\textbf{29.167}} &\textcolor{red}{\textbf{0.794}} &65.367 &3.6069  \\
TADSR &28.387 &0.7758 &\textcolor{red}{\textbf{67.016}} &\textcolor{red}{\textbf{3.7491}}  \\

\hline
\end{tabular}
\label{tab:ablation_blur}
\vspace{-1em}
\end{table}

\section{More Ablation Study}
\subsection{Different Timesteps}
\label{sec:ablation_ts}
\figref{fig:supp_vc_ts} presents TADSR's results at different timesteps $t_s$, demonstrating a gradual transition from fidelity to realism reconstruction as the $t_s$ increases. 
Specifically: (1) In the first row, TADSR progressively generates richer eyelash textures and sharper contours;
(2) The second row shows how patterned shadows gradually transform into stain-like artifacts; 
(3) For the third row, TADSR reconstructs plausible architectural stripes not present in the low-quality input; and
(4) The fourth row reveals emerging yellow pistils in flower centers. 
These progressive changes evidence TADSR's enhanced utilization of the pre-trained generative priors in SD at larger $t_s$, effectively balancing the fidelity-realism trade-off condition on $t_s$.
In addition, we also present the performance of our method at different timesteps.
As shown in~\tabref{tab:ablation_timestep}, with increasing timesteps, reference metrics (PSNR, SSIM~\cite{ssim}) tend to decrease while no-reference metrics (MUSIQ~\cite{ke2021musiq}, TOPIQ~\cite{chen2024topiq}, QALIGN~\cite{qalign-pmlr-v235-wu24ah}) tend to increase. 
This is consistent with the visual comparison results in~\figref{fig:supp_vc_ts}, demonstrating that our method can achieve one-step realism–fidelity controllable generation simply by adjusting the timestep.

\subsection{Blurred MSE Loss}
\label{sec:ablation_mse}
To avoid gradient inconsistency arising from the ill-posed problem of the Real-ISR task while fully leveraging generative prior of SD, we introduce a blurred MSE loss to replace the original MSE loss. Specifically, we first apply a Gaussian blur to both the reconstructed image $G_\theta(x_L)$ and the HQ image $x_H$ before computing the MSE loss. The blurred MSE loss can be formed as:
\begin{equation}
\label{eq:eq13}
    \mathcal{L}_{MSE}^{blur} = \mathcal{L}_{MSE} \left( G_\theta(x_L)*G_{t_s}, x_H*G_{t_s} \right).
\end{equation}
Where $*$ denotes the convolution operation, $G_{t_s}$ is the Gaussian convolution kernel whose size is determined by $t_s$. Let $k_{t_s}$ as the kernel size of $G_{t_s}$, it satisfies:
\begin{equation}
\label{eq:eq14}
    k_{t_s} = 5 + 4*\lfloor \frac{t_s}{200} \rfloor.
\end{equation}
To validate the effectiveness of the proposed blurred MSE loss, we performed an ablation study by removing it. 
As shown in Tab.~\ref{tab:ablation_blur}, when the blurred MSE loss is removed, the no-reference metrics degrade (MUSIQ~\cite{ke2021musiq}, QALIGN~\cite{qalign-pmlr-v235-wu24ah}) significantly while the reference metrics (PSNR, SSIM~\cite{ssim}) improve, demonstrating a trade-off effect where fidelity is enhanced at the expense of realism. 
To better align with the reference metrics, we selected TADSR's output at $t_s=300$. 
With the blurred MSE loss incorporated, TADSR achieves improvements across all metrics, indicating that this loss function enables a more optimal balance between fidelity and realism.

\begin{table}[t]
  \small
  \centering
  \caption{Quantitative comparison of ablation study on the hyperparameters of TAVSD loss, evaluated on RealSR~\cite{realsr} dataset.}
  \setlength\tabcolsep{3.2pt}
    \begin{tabular}{c|c|c|c|c|c} \hline

\multicolumn{2}{c|}{\textbf{Methods}} &
\multirow{2}{*}{PSNR $\uparrow$} &
\multirow{2}{*}{SSIM $\uparrow$}  &
\multirow{2}{*}{MUSIQ $\uparrow$} &
\multirow{2}{*}{MANIQA $\uparrow$} \\
\cline{1-2}
$\lambda$ & $\gamma$ & & & & \\ 
\hline

0.25 & 0   & \textcolor{red}{\textbf{25.757}} & \textcolor{red}{\textbf{0.7218}} & 69.408  & 0.6541  \\
\textbf{0.50} & \textbf{0}   & \textcolor{blue}{\underline{25.166}} & 0.7150 & \textcolor{blue}{\underline{71.182}} & 0.6715  \\
0.75 & 0   & 24.398 & 0.7038 & \textcolor{red}{\textbf{71.484}} & \textcolor{red}{\textbf{0.6774}}  \\
0.50 & 50  & 25.090 & \textcolor{blue}{\underline{0.7159}} & 70.742 & 0.6702  \\
0.50 & 100 & 24.809 & 0.7115 & 70.918 & \textcolor{blue}{\underline{0.6727}}  \\

\hline
\end{tabular}
\label{tab:ablation_hyper}
\vspace{-1em}
\end{table}

\begin{table*}[t]
\fontsize{12pt}{16pt}\selectfont
\centering
\caption{A comprehensive evaluation against state-of-the-art GAN-based methods across synthetic and real-world datasets. The top-performing results under each metric are marked in \textcolor{red}{\textbf{red}}.}
\resizebox{\textwidth}{!}{
\begin{tabular}{@{}c | c | c c c c c c c c @{}}
\hline
Datasets &
Methods &
PSNR $\uparrow$ &
SSIM $\uparrow$  &
LPIPS $\uparrow$  &
CLIPIQA $\uparrow$  &
MUSIQ $\uparrow$  &
MAINIQA $\uparrow$  &
TOPIQ $\uparrow$  &
QALIGN $\uparrow$ \\
\hline
\multirow{4}{*}{\begin{tabular}[c]{@{}c@{}}
\textit{DIV2k-Val}\end{tabular}}
& BSRGAN &\textcolor{red}{\textbf{24.583}} &0.6269 &0.3351 &0.5246 &61.196 &0.5041 &0.5460 & 3.1708  \\
& RealESRGAN &24.293 &\textcolor{red}{\textbf{0.6372}} &0.3112 &0.5277 &61.058 &0.5485 &0.5297 & 3.2768  \\
& LDL &23.828 &0.6344 &0.3256 &0.5179 &60.038 &0.5328 &0.5144 &3.1797  \\
& TADSR &23.815 &0.6028 &\textcolor{red}{\textbf{0.3078}} &\textcolor{red}{\textbf{0.7353}} &\textcolor{red}{\textbf{69.649}} &\textcolor{red}{\textbf{0.6443}} &\textcolor{red}{\textbf{0.7044}} &\textcolor{red}{\textbf{4.0783}}  \\

\hline

\multirow{4}{*}{\begin{tabular}[c]{@{}c@{}}
\textit{DrealSR}\end{tabular}}
& BSRGAN &\textcolor{red}{\textbf{28.701}} &0.8028 &0.2858 &0.5092 &57.165 &0.4845 &0.5060 &2.9580  \\
& RealESRGAN &28.615 &0.8051 &0.2819 &0.4525 &54.268 &0.4903 &0.4623 &2.8645  \\
& LDL &28.197 &\textcolor{red}{\textbf{0.8124}} &\textcolor{red}{\textbf{0.2792}} &0.4475 &53.949 &0.4894 &0.4518 &2.8564  \\
& TADSR &28.387 &0.7758 &0.3235 &\textcolor{red}{\textbf{0.7398}} &\textcolor{red}{\textbf{67.016}} &\textcolor{red}{\textbf{0.6309}} &\textcolor{red}{\textbf{0.6758}} &\textcolor{red}{\textbf{3.7491}}  \\

\hline

\multirow{4}{*}{\begin{tabular}[c]{@{}c@{}}
\textit{RealSR}\end{tabular}}
& BSRGAN &\textcolor{red}{\textbf{26.379}} &\textcolor{red}{\textbf{0.7651}} &\textcolor{red}{\textbf{0.2656}} &0.5116 &63.287 &0.5420 &0.5505 &3.1843  \\
& RealESRGAN &25.686 &0.7614 &0.2710 &0.4494 &60.370 &0.5505 &0.5148 &3.1073  \\
& LDL &25.281 &0.7565 &0.2750 &0.4555 &60.928 &0.5495 &0.5125 &3.0888  \\
& TADSR &25.166 &0.7150 &0.3168 &\textcolor{red}{\textbf{0.7283}} &\textcolor{red}{\textbf{71.182}} &\textcolor{red}{\textbf{0.6715}} &\textcolor{red}{\textbf{0.7082}} &\textcolor{red}{\textbf{3.9477}}  \\

\hline
\end{tabular}
}
\label{tab:compare_with_gan_methods}
\vspace{-2mm}
\end{table*}

\begin{table*}[t]
  \small
  \centering
  \caption{The inference time and the number of parameters of  diffusion-based Real-ISR methods. The top-performing results under each metric are marked in \textcolor{red}{\textbf{red}}.}
  \setlength\tabcolsep{3.2pt}
    \begin{tabular}{c|c|c|c|c|c|c|c|c|c|c} \hline
  &StableSR &DiffBIR & SeeSR & SinSR & OSEDiff & S3Diff & AdcSR & TSDSR & PisaSR & TADSR \\
\hline
 Inference Step & 200 & 50 & 50 & 1 & 1 & 1 & 1 & 1 & 2 & 1 \\
 \hline
 Inference time(s) & 10.40 & 9.83 & 5.64 & 0.1785 & 0.1463 & 0.4704   & \textcolor{red}{\textbf{0.0825}} & 0.0947 & 0.1675 & 0.1465  \\
 \hline
 \#Params(MB) & 1563 & 1682 & 2514 & \textcolor{red}{\textbf{119}} & 1775 & 1327 & 456 & 2207 & 1302 & 1777  \\  

\hline
\end{tabular}
\label{tab:efficency}
\vspace{-1em}
\end{table*}

\subsection{Hyperparameters in TAVSD}
\label{sec:ablation_hyper}
To verify the sensitivity of our method to the hyperparameters in TAVSD, we conducted ablation studies by varying their values.
We set the timestep $t_s$ to 500 and evaluated our method under different hyperparameters on the RealSR~\cite{realsr} dataset. 
As shown in~\tabref{tab:ablation_hyper}, with the increase of $\lambda$ and $\gamma$, our method exhibits a decrease in reference metrics while no-reference metrics improve, reflecting a trade-off of fidelity for enhanced realism. 

This phenomenon is consistent with the functionality of TAVSD. 
As discussed in~\secref{sec:intro}, the pre-trained SD model exhibits different generative priors at different timesteps: smaller timesteps tend to favor fidelity, while larger timesteps tend to favor generation. 
This also means that the teacher model in VSD will provide generation guidance based on semantic priors when the time step is large.
Therefor, when $\lambda$ and $\gamma$ increase, the same $t_s$ is mapped to a larger $t_v$, causing the teacher model to provide guidance more biased toward generation. 
In terms of metrics, this is manifested as an increase in no-reference metrics and a decrease in reference-based metrics.
Considering the balance between realism and fidelity, we ultimately choose $\lambda=0.5$ and $\gamma=0$ as the default setting for our model.

\section{Comparison of Efficiency with Other One-step Real-ISR Methods}
\label{sec:efficiency}
We compare the number of parameters and inference time of one-step diffusion-based Real-ISR models in~\tabref{tab:efficency}.
Inference time is measured on the $\times$4 SR task with 128 $\times$ 128 LQ images using a single NVIDIA 3090 24G GPU.
Compared with OSEDiff~\cite{wu2024one}, our method achieves roughly the same inference time and parameter count, while showing significant improvements in no-reference metrics and visual quality. 
PisaSR requires two inferences to achieve controllable Real-ISR due to the presence of two LoRA weights. 
In contrast, our method can obtain controllable Real-ISR with a single inference simply by adjusting the time step, resulting in fewer inference steps and shorter inference time.

\begin{figure*}[ht!]
  \includegraphics[width=\linewidth]{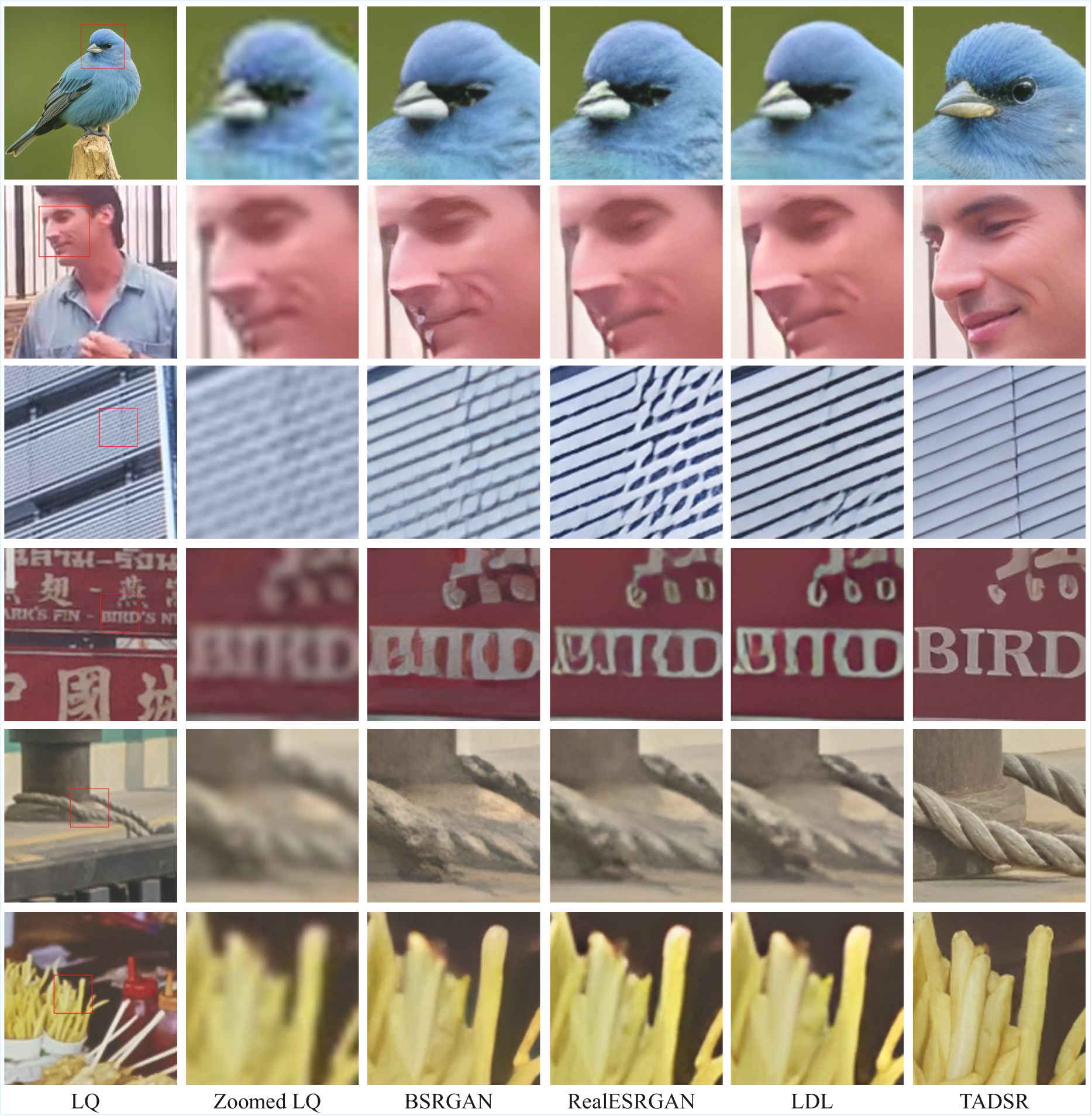}
  \caption{Vision comparisons between TADSR and GAN-based Real-ISR methods. Zoom in for a better view.}
\label{fig:supp_vc_gan}
\vspace{-1em}
\end{figure*}
\section{Comparisons with GAN-based Real-ISR Methods}
\label{sec:compare_gan}
We compare TADSR with three GAN-based Real-ISR methods: BSRGAN~\cite{zhang2021designing}, RealESRGAN~\cite{wang2021real}, and LDL~\cite{liang2022details}. 
Quantitative evaluations are conducted on the DIV2K~\cite{div2k}, RealSR~\cite{realsr}, and DRealSR~\cite{drealsr} datasets, with results summarized in Tab.~\ref{tab:compare_with_gan_methods}. 
The experimental results demonstrate that TADSR, leveraging the powerful generative priors of the pre-trained SD, achieves significantly superior no-reference metrics (e.g., CLIPIQA~\cite{clipiqa}, MAINIQA~\cite{yang2022maniqa}) compared to GAN-based methods.

Additionally, Fig.~\ref{fig:supp_vc_gan} presents a visual comparison between TADSR and other GAN-based methods. 
The results show that TADSR reconstructs more photorealistic and natural outcomes, including higher fidelity in text and architectural structures (from the first to the third group), and more realistic rope textures (in the fourth group).

\section{More Visual Comparisons with SD-based Real-ISR Methods}
\label{sec:vc_sd}
\begin{figure*}[ht]
  \includegraphics[width=\linewidth]{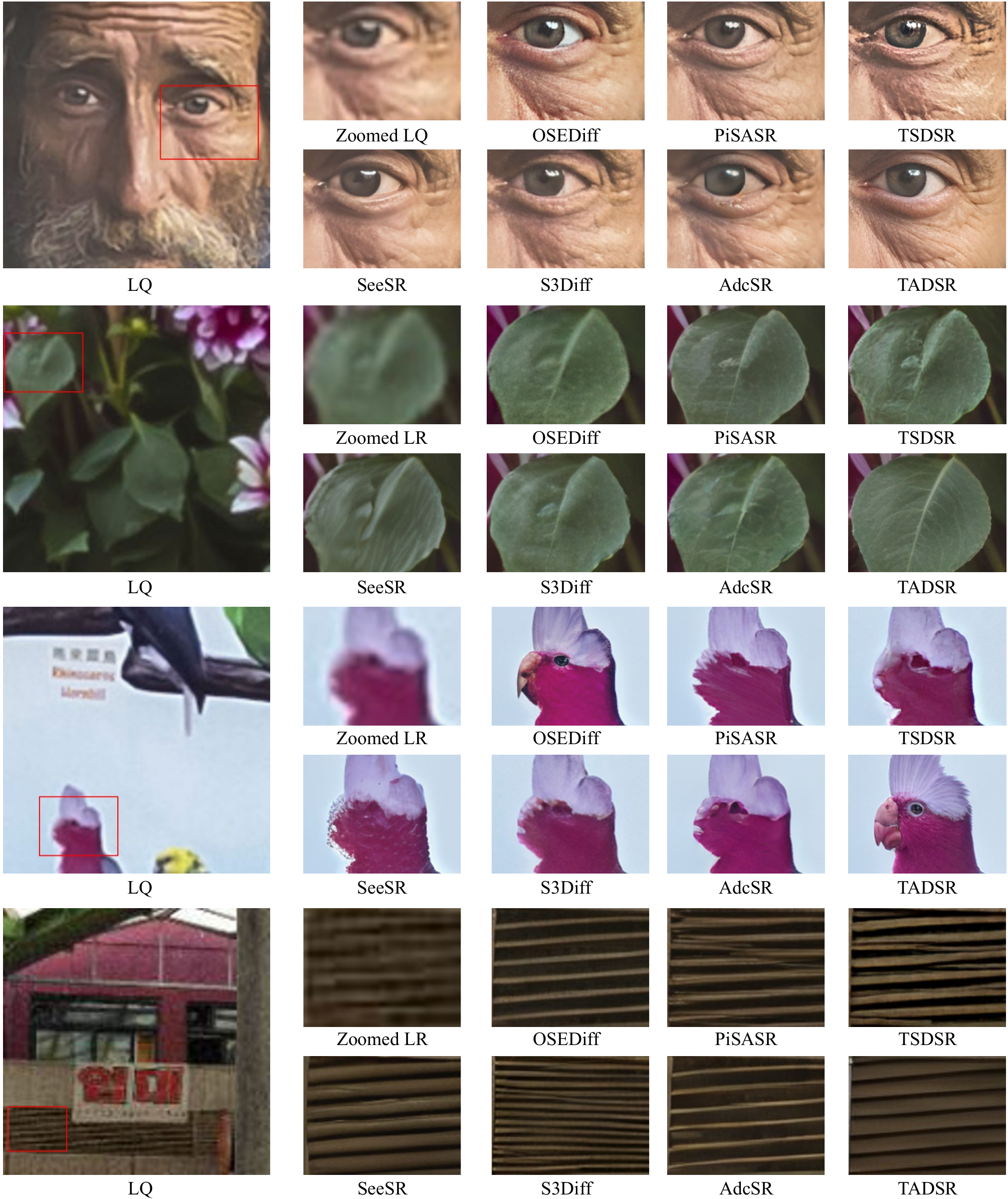}
  \caption{Vision comparisons between TADSR and SD-based Real-ISR methods (SeeSR~\cite{wu2023seesr}, OSEDiff~\cite{wu2024one}, S3Diff~\cite{zhang2024degradation}, PiSASR~\cite{sun2025pixel}, AdcSR~\cite{chen2025adversarial}, TSDSR~\cite{dong2025tsd}). Zoom in for a better view.}
\label{fig:supp_vc_1}
\end{figure*}

\begin{figure*}[ht]
  \includegraphics[width=\linewidth]{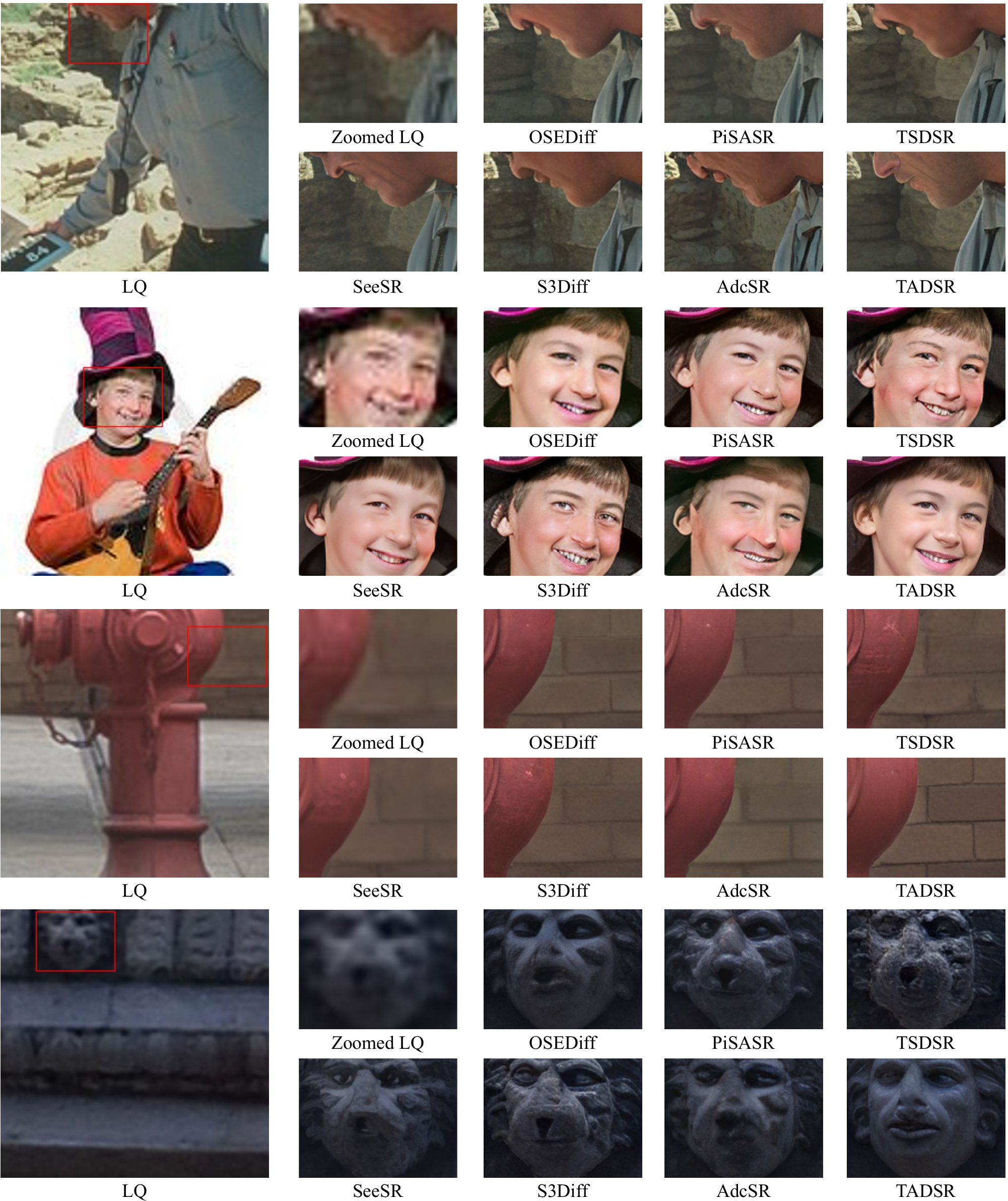}
  \caption{Vision comparisons between TADSR and SD-based Real-ISR methods (SeeSR~\cite{wu2023seesr}, OSEDiff~\cite{wu2024one}, S3Diff~\cite{zhang2024degradation}, PiSASR~\cite{sun2025pixel}, AdcSR~\cite{chen2025adversarial}, TSDSR~\cite{dong2025tsd}). Zoom in for a better view.}
\label{fig:supp_vc_2}
\end{figure*}

\begin{figure*}[ht]
  \includegraphics[width=\linewidth]{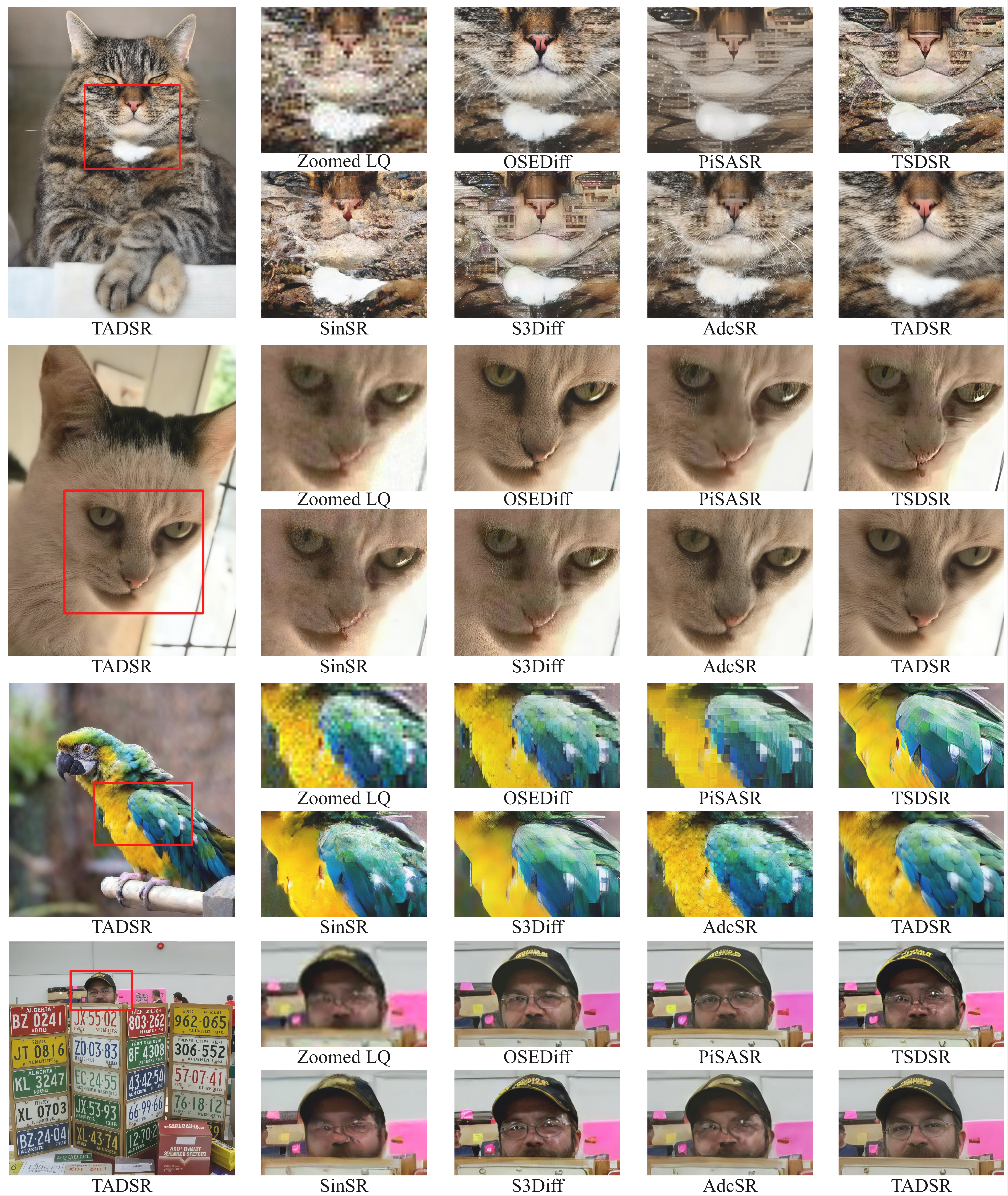}
  \caption{Vision comparisons between TADSR and Diffusion-based Real-ISR methods (SinSR~\cite{wang2023sinsr}, OSEDiff~\cite{wu2024one}, S3Diff~\cite{zhang2024degradation}, PiSASR~\cite{sun2025pixel}, AdcSR~\cite{chen2025adversarial}, TSDSR~\cite{dong2025tsd}). Zoom in for a better view.}
\label{fig:supp_vc_3}
\end{figure*}

We provide more visual comparisons between TADSR and other SD-based SR methods in Fig.~\ref{fig:supp_vc_1} and Fig.~\ref{fig:supp_vc_2}. Compared to other methods, TADSR consistently produces clearer, more realistic, and more natural results. 
Moreover, although our training is conducted at a resolution of 512×512, we provide visual comparisons of TADSR and other diffusion-based one-step Real-ISR methods on 2K-resolution images. 
As shown in~\figref{fig:supp_vc_3}, TADSR is also capable of maintaining strong structural consistency and producing realistic, natural SR results on high-resolution images. 
Under severe synthetic degradations (first and third rows), TADSR still demonstrates powerful deblurring capability and highly realistic generative effects, whereas other single-step SR methods are noticeably affected by the degradations, exhibiting clear artifacts or color smearing. 
In real-world degradation scenarios (second and fourth rows), TADSR similarly recovers more authentic texture details (e.g., the cat’s fur and nose) and more natural structures (e.g., the person’s eyes and glasses).
These visual comparisons consistently demonstrate that TADSR makes effective and thorough use of the SD generative prior.

\end{document}